\documentclass[showpacs,prl,onecolumn,aps,superscriptaddress,preprintnumbers,nofootinbib]{revtex4}
\usepackage[T1]{fontenc}
\usepackage[latin9]{inputenc}
\usepackage{amsmath,amssymb,bigints}
\usepackage{epsfig}
\usepackage{graphicx}
\usepackage{amsmath}
\usepackage{amsfonts}
\usepackage{epstopdf}
\def\slashchar#1{\setbox0=\hbox{$#1$}     		
   \dimen0=\wd0                                 	
   \setbox1=\hbox{/} \dimen1=\wd1               	
   \ifdim\dimen0>\dimen1                        	
      \rlap{\hbox to \dimen0{\hfil/\hfil}}      	
      #1                                        	
   \else                                        	
      \rlap{\hbox to \dimen1{\hfil$#1$\hfil}}   	
      /                                         	
   \fi}

\renewcommand{\vec}{\boldsymbol}
\newcommand{\beq}{\begin{equation}}
\newcommand{\eeq}{\end{equation}}
\newcommand{\bea}{\begin{eqnarray}}
\newcommand{\eea}{\end{eqnarray}}
\newcommand{\baa}{\begin{array}}
\newcommand{\eaa}{\end{array}}

\def\eq#1{{Eq.~(\ref{#1})}}
\def\fig#1{{Fig.~\ref{#1}}}

\newcommand{\bas}{\bar{\alpha}_S}
\newcommand{\as}{\alpha_S}

\newcommand{\nn}{\nonumber}

\newcommand{\h}{\frac{1}{2}}

\newcommand{\Lb}{\left(}
\newcommand{\Rb}{\right)}

\renewcommand{\vec}[1]{\boldsymbol{#1}}

\begin{document}
\title{ Thermal radiation  and inclusive production in  the KLN model for ion-ion
 collisions.}
\author{ E.~ Gotsman,}
\email{gotsman@post.tau.ac.il}
\affiliation{Department of Particle Physics, School of Physics and Astronomy,
Raymond and Beverly Sackler
 Faculty of Exact Science, Tel Aviv University, Tel Aviv, 69978, Israel}
 \author{ E.~ Levin}
 \email{leving@tauex.tau.ac.il, eugeny.levin@usm.cl} \affiliation{Department of Particle Physics, School of Physics and Astronomy,
Raymond and Beverly Sackler
 Faculty of Exact Science, Tel Aviv University, Tel Aviv, 69978, Israel} 
 \affiliation{Departemento de F\'isica, Universidad T\'ecnica Federico Santa Mar\'ia, and Centro Cient\'ifico-\\
Tecnol\'ogico de Valpara\'iso, Avda. Espana 1680, Casilla 110-V, Valpara\'iso, Chile} 

\date{\today}


\begin{abstract}
  
     We show that  in order to 
obtain a successful description of
the transverse
momenta distribution for charged particles in ion-ion collisions, 
one must include a thermal emission term.  
 The temperature of this emission $T_{\rm th}$  turns out to be
 proportional to the saturation scale, $T_{\rm th} = 1.8/2\pi  \,Q_s$.    
  The formalism  for the calculation of the transverse
 momenta spectra in CGC/saturation approach  is developed, in which
 two stages of the process are seen: creation of  the colour 
glass
 condensate, and  hadronization of the gluon jets.
 Our calculations  are based on the
 observation that even for small values of $p_T$, the main contribution in
 the integration over the dipole sizes 
 stems from the kinematic region in vicinity of the saturation momentum, 
where theoretically, we know the scattering amplitude.  Non-perturbative
 corrections  need to be included in the model
 of hadronization. This model incorporates  the decay of a gluon jet with 
 effective mass  $m^2_{\rm eff} = 2 Q_s \mu_{\rm soft}$ where $\mu_{\rm
 soft}$ denotes the soft scale,  with the fragmentation functions 
 at all  values of the transverse momenta.

  We use  the KLN model  which,   provides a  simple way to estimate 
the 
 cross sections for the different centrality classes.
Comparing the results of this paper with  the transverse distribution in
 the proton-proton scattering,  we see two major differences. First, a
 larger contribution of the thermal radiation term is needed,  in accord 
 with higher parton densities of  the  produced colour glass condensate. 
Second,  even changing the model for the hadronization,
  without a thermal radiation term,
we fail to describe the $p_T$ spectrum.
  Consequently, we   conjecture  that the
 existence of the thermal radiation term  is independent of the model 
of
 confinement.

\end{abstract}

\preprint{TAUP - 3030/18}

\maketitle

\tableofcontents

\flushbottom

\section{Introduction}

In this paper we continue to discuss the  processes  of multi-particle
 generation at high energy, in the framework of the Color Glass
 Condensate(CGC)/saturation approach (see Ref.\cite{KOLEB} for the review). 
The main ideas of our approach  have  been discussed in our paper
 (see Ref.\cite{GOLETE}) for   hadron-hadron scattering at
 high energies. Here,  we  consider
   heavy ion collisions  in which, we believe, the distinctive features
 of the CGC/saturation approach manifest themselves in the clearest way.
 For theoretical descriptions of these processes we have to develop a
 non-perturbative approach, since these processes occur at long distances.
 In particular, we have to deal with the unsolved problem of the confinement
 of quarks and gluons. Fortunately, in the framework of the CGC/saturation
 approach, the most basic features of the processes of the multi-particle
 generation stem from the production of the  new phase of QCD: the dense
 system of partons (gluons and
 quarks) with a new characteristic
 scale: saturation momentum $Q_s(W)$,  which increases as a 
 function of energy $W$\cite{GLR,MUQI,MV}.  In ion-ion collisions the
 new phase of QCD is produced with a larger density than in hadron-hadron
 scattering, and we believe,  that the  essential properties of this 
phase
 will  manifest themselves in a clear way. However, 
  the  transition
 from this system of partons to the measured state of hadrons, is still
 an unsolved problem.
 
  Due to our lack of
 theoretical understanding of the confinement of quarks and gluons,
at the moment, we need to use a 
  pure phenomenological input for 
the long distance non-perturbative physics.
  In
 particular,  we wish to use phenomenological fragmentation
 functions.  Hence, our model for confinement is that the parton
 (quark or gluon)  with the transverse momenta of the order of $Q_s$
 decays into  hadrons  according to the given fragmentation 
functions. 
 Experimental data  supports this model of hadronization, which  
 provides the
 foundation of all Monte Carlo simulation programs, and  leads to
 descriptions of the 
 transverse momenta distribution of the hadrons, at the LHC energies.
  As an example, we refer to Ref.\cite{ALICE13}, which shows that 
 next-to-leading order QCD calculations with formation of the hadrons 
 in accord with the fragmentation functions (\cite{NLOFIT}),  is able
 to describe the transverse momentum spectra, for the LHC range of
 energies. In a sense, at present, this model is the best that we 
can  propose
  to describe   multi - hadron production\footnote{We have to mention,
 that actually the model of hadronization with given fragmentation function
  describes the experimental data at large values of $p_T$
 ($p_T \,\geq\,3\,GeV$ \cite{NLOFIT}  and  $p_T \,\geq\,5\,GeV$
 \cite{NLOFIT1}).  In our model of the hadronization we assume
 that we can describe the data  with the fragmentation function
 at any value of $p_T$. We   will  demonstrate that
   such a description is possible.}.

  The space-time picture of the high energy interaction in the 
CGC/saturation approach,  is as follows:
The parton 
configuration in QCD is formed long before the interaction at
 distances $R_A/ x$, where $R_A$ - denotes the nucleus radius, and $x$ 
 the fraction of longitudinal momentum carried by the parton which
 interacts with the target. However, before the collision, the wave
 function of this partonic fluctuation is the eigenfunction of
 the Hamiltonian  and, therefore, the system has  zero entropy. 
The interaction with the target of  size $R_A$ destroys the coherence
 of the parton wave
function of the projectile. The typical time, which is needed for this,
 is of the order of
$\Delta t \propto R_A$, and is much smaller than the lifetime of all 
faster partons in the fluctuation. Hence, this interaction  can be
 viewed as a rapid quench of the
entangled partonic state\cite{KHLE} with substantial  entanglement
  entropy.  After this rapid quench, the interaction of the gluons 
 change the Hamiltonian. Since in the CGC/saturation approach
all partons with rapidity larger than that of a particular gluon $y_i$, 
  live longer than this parton,  so  they can be considered  to 
be the  source
 of the classical
field that emits this gluon.  It  has been  shown that  after the 
quench,
 the fast gluons create the  longitudinal chromo-electrical 
background field. Moving in this field the gluon accelerates and emits gluons
 which have  the thermal distribution ( the first term of \eq{SUM} below)
\cite{KHTU,KLT,BAKH}.
 The temperature of this distribution is intimately related to
 the saturation momentum, which provides  the only dimensional 
scale in the colour glass condensate. It determines both the strength
 of the longitudinal fields and the  ultraviolet cutoff on the
 quantum modes, resolved by the collision.  It turns out
 \cite{KHTU,KLT,BAKH,DUHA,CKS} that
 \beq \label{THTEM}
  T_{\rm th}\,\,=\,\,{\rm c} \frac{Q_s}{2 \pi}
  \eeq
  with the semi-classical estimates 
  \cite{KLT} for the constant $c = 1.2$.
  
  The appearance
 of a thermal emission in  a high energy proton-proton collision is a
 remarkable feature of the interaction, since the number of the
 secondary interactions in proton-proton collisions is rather
 low, and cannot provide the thermalization due to the interaction 
in the final state.  The origin of  thermal radiation in the
 framework of the CGC approach was clarified a decade ago
 \cite{KHTU,DUHA,CKS,KLT} and, recently, the new idea that
 the quantum entanglement is at the origin of the parton densities,
 has been added to these arguments\cite{KHLE}. 
   
  The goal of this paper is to re-visit  the process of  inclusive 
production in the 
 CGC/saturation approach, for   more thorough 
consideration,  and to show that the thermal term with the temperature
 given by \eq{THTEM}, is needed  for describing  the experimental
 data for ion-ion collisions at high energies.  At 
first sight, it 
 looks as we are   pushing  at  an open door, since 
  it has been shown \cite{FIT0,FIT1,FIT2,FIT3,BAKH,FPV}  that the 
 experimental
 data \cite{ALICE13,ALICE2,ATLAS1,ATLAS2,CMS1,CMS2} at high energy
 both for hadron-hadron and ion-ion scattering , can
 be described as the sum of two terms:
\beq \label{SUM}
\frac{d \sigma}{dy d^2 p_T}\,\,=\,\,\underbrace{A_{\rm therm} e^{- \frac{m_T}{T_{\rm th}}}}_{\mbox{ thermal radiation}}\,\,\,\,+\,\,\,\,\underbrace{A_{\rm hard}\frac{1}{\Lb 1 + \frac{m^2_T}{T^2_{\rm h}\,n}\Rb^n}}_{\mbox{hard emission}}
\eeq
with
\beq \label{T} 
 T_{\rm th}\,\,=\,\,0.098 \Lb \sqrt{\frac{s}{s_0}}\Rb^{0.06} \,{\rm GeV};~~~~~~~~~~
  T_{\rm h}\,\,=\,\,0.409 \Lb \sqrt{\frac{s}{s_0}}\Rb^{0.06} \,{\rm GeV}; 
  \eeq  
  
  The first term in  \eq{SUM} is the desired thermal radiation, while
 the second term  describes  hadron production in the hard processes,
 showing the power-like decrease at high $p_T$. The same dependence on
 energy of both $T_{\rm th}$ and  $ T_{\rm h} $ supports the  main
 idea of CGC, i.e.  both parameters are related to the saturation 
momentum 
($Q_s$).
 However, it turns out that this dependence  differs from the saturation
 scale $Q_s \propto  \Lb \sqrt{\frac{s}{s_0}}\Rb^{\lambda}  $.

  The value of $\lambda $ can be calculated theoretically and measured
 experimentally. The leading order QCD evaluation leads to
 $\lambda = 4.9\bas$, where $\bas$ denotes the running QCD coupling.
 Plugging in a reasonable estimate for $\bas\Lb Q_s\Rb \approx 0.2$,
  $\lambda$ turns out to be large, about 0.8-1. The
 phenomenological description of the hard processes both for nucleus
 interactions\cite{DKLN} and DIS( see Ref.\cite{RESH} and references
 therein), give the value of $\lambda = 0.2-0.24$. Thus, 
  in the CGC approach we expect  $$T_{\rm th} \propto\,\,T_{\rm h}
 \propto \Lb \sqrt{\frac{s}{s_0}}\Rb^{\lambda/2} \sim \Lb \sqrt{\frac{s}{s_0}}
\Rb^{0.1 - 0.112},  $$ 

Hence, in spite of the fact that
\eq{SUM} and \eq{T} show that both temperatures 
have dependence on energy in accord with the CGC result, this dependence
  contradicts the CGC prediction. Especially, the different
   $Q_s$ energy dependence 
  in  the second term of  \eq{SUM}, which
 corresponds to the 
  contribution of the hard processes,  looks strange if not wrong, since  
 the CGC approach to the hard processes has been confirmed both
 theoretically and experimentally,  and  we know that the typical
 scale in these processes,  is the saturation momentum. The second
 remark is related to the value of the hard contribution. In the CGC
 approach  it should be calculated  theoretically,  and  not  be 
determined by
 a  fitting procedure.

\section{Inclusive production in CGC/saturation approach for ion-ion collison}

\subsection{General formulae}

The general formula for  the gluon jet production in ion-ion collisions
 in the CGC/saturation approach,
 has the following  form 
 (see Ref.\cite{KTINC} for the proof):

\bea \label{MF}
\frac{d \sigma_G}{d y \,d^2 p_{T}}\,\,& =\,\,& \frac{2C_F}{\alpha_s (2\pi)^4}\,\frac{1}{p^2_T} \,\int d^2  r\,e^{i \vec{p}_T\cdot \vec{r}}\,\,\,\int d^2 b\,\nabla^2_T N^{A_1}_G\Lb y_1 = \ln(1/x_1); r, b \Rb\,\,\int d^2 b' \,\nabla^2_T\,N^{A_2}_G\Lb y_2 = \ln(1/x_2); r,  b' \Rb. 
\eea
where $N^{A_i}_G\Lb y_1 = \ln(1/x_1); r, b \Rb$ can be found from
 the amplitude
 of the dipole-nucleus  scattering 
$N^{A_i}\Lb y_i = \ln(1/x_i); r_; b \Rb$ :
\beq \label{NG}
N^{A_i}_G\Lb y_i = \ln(1/x_i); r,  b \Rb\,\,=\,\,2 \,N^{A_i}\Lb y_i = \ln(1/x_i); r, b \Rb\,\,\,-\,\,\,\Lb N^{A_i}\Lb y_i = \ln(1/x_i); r,  b \Rb\Rb^2
\eeq
where $r$ denotes the size of the dipole, $b$  it's impact parameter 
and 
\beq \label{X}
x_1 \,\,=\,\,\frac{p_T}{W} \,e^{y};\,\,\,\,\,x_2 \,\,=\,\,\frac{p_T}{W}
 \,e^{-y};\eeq

where $y$ denotes the rapidity of the produced gluon in c.m.f. and $W$ the
 c.m.s. energy of the collision.  In this paper we consider the gluon
 production at $y = 0$.
$C_F = (N^2_c - 1)/2 N_c$ and $  \bas \,= \,\as N_c/\pi$ with the number
 of colours equals $N_c$. $\as$ denotes  the running QCD coupling, 
$\nabla^2_T$ 
  the Laplace operator with respect to $r$,  it is equal to 
$\nabla^2_T\,=\,\frac{1}{r} \frac{d}{d r} \Lb r \frac{d}{d r}\Rb$.

In our paper\cite{GOLETE} we found that the main contribution at high
 energies stem from the specific kinematic region in the vicinity of
 the saturation scale. Indeed, at high energies and sufficiently small
 values of $p_T$
 the dipole amplitudes
 are in the saturation region, where the parton densities are large  and 
the
 dipole scattering amplitude displays  geometric scaling behaviour, being
 a function of  only one variable:  $ \tau \,=\,r \,Q_s\Lb W, b\Rb$ \cite{GS1,GS2,GS3}.  Deep in the saturation region the dipole amplitude  tends to approach 1, but $\nabla^2_T N^{A_1}_G\Lb y_1 = \ln(1/x_1); r, b \Rb \,\,\xrightarrow{\tau \,\gg\,1} \,\,0$. Consequently the main contribution in \eq{MF} stems from the kinematic region where $\tau \sim 1$ or, in other words, from the vicinity of the saturation scale. The most attractive features of this observation is the fact that we know theoretically the behaviour of the scattering amplitude in this region. It has the form:
 \beq \label{VQS}
 N\Lb y_i = \ln(1/x_i); r, b \Rb \,\,=\,\,\,N_0\, \Lb r^2\,Q_s\Lb A,Y, b\Rb\Rb^{\bar{\gamma}}\,\,=\,\,N_0\, \Lb\tau\Rb^{
2\,\bar{\gamma}} 
 \eeq
 where $\bar{\gamma} \,=\,1 - \gamma_{cr}$ and $\gamma_{cr} \,=\,0.37$
 in the leading order is the solution to the equation\cite{KOLEB}:
\beq \label{GACR}
\frac{d \chi\Lb \gamma_{cr}\Rb}{d \gamma_{cr}}\,\,=\,\,-\,\frac{\chi\Lb \gamma_{cr}\Rb}{ 1\,-\,\gamma_{cr}}.
\eeq
$\chi\Lb \gamma\Rb$ is given by  
 \beq \label{KER}
 \chi\Lb \gamma\Rb\,\,=\,\,2 \psi(1)\,-\,\psi(\gamma))\,-\,
\psi(1 - \gamma)
\eeq 
Note that $\omega = \bas \chi\Lb \gamma\Rb$ is the eigenvalue of the BFKL
 equation\cite{BFKL}.

The saturation momentum $Q_s\Lb A, Y, b\Rb $  has the following dependence
 on energy\cite{GLR,MUQI,MV}
\beq \label{QS}
Q_s\Lb A, Y, b\Rb\,\,=\,\,Q_s\Lb A, Y = Y_0 , b\Rb\,e^{\lambda \,\Lb Y - Y_0\Rb}
\eeq
where $\lambda = \bas \chi\Lb \gamma_{cr}\Rb/(1 - \gamma_{cr})$ in the LO
 of perturbative QCD.
We will discuss the dependence of the saturation momentum on A below. It
 should be stressed that \eq{VQS} together with \eq{QS} give the correct
 behaviour $N \propto \exp\Lb - \mu b\Rb$ of the scattering amplitude
 at large impact parameters, which is determined by the non-perturbative
 behaviour of the saturation momentum at the initial energy ($Y_0$). It
 should be compared with the general case, for which at the moment we are
 not able to modify the main equations in a such way to obtain this
 behaviour \cite{KOWI}. Note, that we can find $N_G$ of \eq{NG} only
 if we know the impact parameter behaviour of the scattering amplitude.

Hence, we  conclude that $\nabla^2_T N^{A_1}_G\Lb y_1 = \ln(1/x_1);
 r, b \Rb$ and  $\nabla^2_T N^{A_1}_G\Lb y_2 = \ln(1/x_2); r, b \Rb$ in
 \eq{MF}, as well as the integral over $r$, can be calculated in the
 framework of CGC/saturation approach, and there is no need to introduce
 the non-perturbative corrections due to the unknown physics at long
 distances (see Refs.\cite{WAZW,KHALE} for example) in the dipole
 scattering amplitude. However, plugging  the scattering
 amplitude of \eq{VQS} in \eq{MF}, one can see  that $\int d^2  r\,e^{i 
\vec{p}_T\cdot
 \vec{r}}\,\,\,\int d^2 b\,\nabla^2_T N^{A_1}_G\,\,\int d^2 b'
 \,\nabla^2_T\,N^{A_2}_G$   is not suppressed at $p_T \to 0$. Therefore,
 we have $\frac{d \sigma_G}{d y \,d^2 p_{T}}\,\propto\,1/p^2_T$.
 This divergence can only be tamed  in the process of hadronization, 
which
 has to be treated using a non-perturbative approach to QCD.

At the moment, as has been alluded to in the introduction, we can treat
 the hadronization only using phenomenological models. Our model consists
 of two elements. First, every gluon  decays into the jet of hadrons
 with a known fragmentation function:
 \beq \label{XSPION}
 \frac{d \sigma^{\pi}}{d y \,d^2 p_{T}} \,\,=\,\,\int^1_0  d x_G\,
 \frac{d \sigma_G}{d y \,d^2 p_{T}} \Lb \frac{p_T}{x_\pi}\Rb\,D^\pi_G\Lb x_\pi\Rb
 \eeq
 We take the fragmentation function $D^\pi_G$  from Ref.\cite{FRFU} 
 which has
 the form
 \beq \label{FRF}
 D^\pi_G\Lb x^\pi\Rb\,\,=\,\,2.17 z^{\alpha} (1-z)^{\beta}
 \left(20 (1-z)^{\gamma1}+1\right); 
 \eeq 
 
 with $\alpha = 0.899$,   $\beta= 1.57$ and $\gamma=4.91$.
 It should be stressed, that we assume that \eq{XSPION} holds for any
 value of the transverse momenta of gluons, while  it is shown that 
 this equation gives a good description of the experimental data for 
rather large
 transverse momenta: $p_T\, > \,3\,GeV$ \cite{NLOFIT}
 and $p_T \,>\,5\,GeV$\cite{NLOFIT1}.

 We note that \eq{MF} as well as \eq{XSPION} lead to a cross
 section which is proportional to $1/p^2_T$. This behaviour results in a
 logarithmic divergency of the integral over $p_T$, or in other words
 gives an infinite number of produced pions at fixed rapidity.
  The reason for this problem, is that we neglected the mass of the
 jet of hadrons that stems from the decay of the gluon. The simple
 estimates \cite{KLN2} give  for a gluon with the value of the
 transverse momentum $p_T$,  the mass of the jet $m^2_{\rm jet}\,=\,2 
p_T\, m_{\rm eff}$, where $
  m_{\rm eff} = \sqrt{m^2 + k_T^2 + k^2_L} - k_L$, $m$ is the mass 
of
 the lightest hadron in the jet, $k_T$ is it's transverse momentum 
and $k_L
 \approx \,k_T$ is the longitudinal momentum of this hadron.
 Since  most pions stem from the decay of $\rho$-resonances we expect
 that $ m_{\rm eff} \approx m_\rho$.
 
 As we have mentioned  in the introduction,   our model for 
 confinement is that  of  the CGC approach, the typical momentum for the
 produced gluon is the saturation momentum. Hence,  most hadrons are
 created  in the jets with the mass $m^2_{\rm jet} \,=\,2 Q_s\,m_{\rm eff}$.
  However, for rare gluons with $p_T \gg  Q_s$ we still have $m^2_{\rm jet}
 \,=\,2p_T\,m_{\rm eff}$ . For  numerical  estimates we
 use $m^2_{\rm jet}\,=\,2 \Lb Q_s \Theta(  Q_s  -  p_T) + p_T 
\Theta( p_T  -  Q_s) \Rb m_{\rm eff}$ which has these two limits.
 $\Theta(x)$ denotes the step function. Using the same idea we replace 
\eq{X} by
 
 \beq \label{XF}
 x \,\,=\,\,\frac{p_T}{W} \,\,=\,\,\frac{ Q_s \Theta(  Q_s  -  p_T)
 + p_T \Theta( p_T  -  Q_s) }{W}. 
 \eeq
 
 Concluding,  we see that our model of the hadronization processes
 includes two ingredients: the fragmentation function of \eq{FRF}
 which gives us the number and $p_T$ distribution of the produced
 hadron from one gluon; and the replacement $1/p^2_T $ in \eq{MF}
 by $1/\Lb p^2_T \,+\,m^2_{\rm jet}\Rb$ with $m^2_{\rm jet}
 \,=\,2\Lb Q_s \Theta(  Q_s  -  p_T) + p_T 
\Theta( p_T  -  Q_s)\Rb \,m_{\rm eff} $.

\subsection{Dipole-nucleon scattering amplitude}
For  completeness of presentation, in this subsection we give a brief
 review of our theoretical and phenomenological approach to the 
contribution
 of the dipole-nucleon scattering amplitude to the inclusive gluon
 production, that has been discussed  in Ref.\cite{GOLETE}. 

\begin{boldmath}
\subsubsection{$ \tau \sim 1.$}
\end{boldmath} In the vicinity of the saturation scale the
 dipole-nucleon 
scattering amplitude has a general form of \eq{VQS} which leads to
\beq \label{DNA1}
\nabla^2_T N_G\Lb Y; r, b \Rb\,\,=\,\,\frac{8 \,\bar{ \gamma} ^2\, N_0 (r\,Q_s)^{2 \bar{\gamma} }}{r^2}  \left( 1\,-\,2 \,N_0\,(r\, Q_s)^{2 \bar{\gamma }}\right)\,\,=\,\,\frac{8\,\bar{\gamma} ^2 \,N_0 \tau^{2 \bar{\gamma} }}{r^2}  \left( 1\,-\,2 \,N_0\,\tau^{2 \bar{\gamma }}\right)\eeq 

\begin{boldmath}
\subsubsection{$ \tau \,>\, 1$. } 
\end{boldmath} Inside of the saturation domain we suggest
  using  $N \,=\,1 - \exp\Lb - \phi_0 \tau^{2 \bar{\gamma}}\Rb$ which 
gives
\bea \label{DNA2}
\nabla^2_T N_G\Lb Y; r, b \Rb\,\,&=& 
\frac{8\, \bar{ \gamma} ^2\, \phi_0 \,(r\,Q_s)^{2 \bar{\gamma} }}{r^2}  \left( 1\,-\,2 \,\phi_0\,(r\, Q_s)^{2 \bar{\gamma }}\right)\exp\Lb- 2 \,\phi_0\, (r\,Q_s)^{2 \bar{\gamma} }\Rb\nn\\
&\,\,=\,\,&\,\,\frac{8 \,\bar{\gamma} ^2\, \phi_0 \tau^{2 \bar{\gamma} }}{r^2}  \left( 1\,-\,2 \,\phi_0\,\tau^{2 \bar{\gamma }}\right) \,e^{ - 2\, \phi_0\,\tau^{2 \bar{\gamma}}}
\eea
We need to match this formula with \eq{DNA1} at $\tau=1$. 
 For $N_0 \ll 1$ 
we obtain  $\phi_0 = N_0$. 

 Ref.\cite{GOLETE},  checked that \eq{DNA2} describes {\bf to a}
 good accuracy the exact solution for the non-linear Balitsky-Kovchegov
 \cite{BK} equation, for the  leading twist BFKL kernel. We wish to remind 
the reader  that 
the scattering amplitude in this kinematic region only gives   a small
 contribution (not more than 15\%) event at $p_T=0$.

\begin{boldmath}
\subsubsection{ $ \tau \,<\, 1$.}   
\end{boldmath} 

In this region we can safely use the perturbative QCD approach
 for the 
scattering amplitude. Therefore, we need to solve the BFKL evolution
 equation in this region which has the following form:

  \beq \label{BFKL}
\displaystyle{\frac{\partial N\Lb Y; \vec{x}_{01}, \vec{b}
 \Rb}{\partial Y}}\,=\,\displaystyle{\frac{\bas}{2\,\pi}\int d^2 
 \mathbf{x_{2}}\,K\Lb \vec{x}_{01};\vec{x}_{02},\vec{x}_{12}\Rb\Bigg\{
 2\, N \Lb Y; \vec{x}_{02},\vec{b} - \h \vec{x}_{12}\Rb \,-\,N\Lb Y; \vec{x}_{01
},\vec{b} \Rb}\Bigg\}
\eeq
where
\beq \label{K}
 \displaystyle{K\Lb \vec{x}_{01};\vec{x}_{02},
\vec{x}_{12}\Rb}\,\,=\,\,  \displaystyle{\frac{\mathbf{x^2_{01}}}{
\mathbf{x^2_{02}}\,\,
{\mathbf{x^2_{12}}} } }
\eeq
 $N\Lb Y; \vec{x}_{01}, \vec{b}\Rb$ denotes the dipole scattering 
amplitude. $\vec{x}_{01} = \vec{x}_1 - \vec{x}_0 \equiv \vec{r}$  the
 size of the dipole. The kernel $K\Lb \vec{x}_{01};\vec{x}_{02},
\vec{x}_{12}\Rb$ describes the decay of the dipole with size $x_{01}$
 into two dipoles of  size: $\vec{x}_{02}$ and $\vec{x}_{12} = 
\vec{x}_{01} - \vec{x}_{02}$.  After  integrating \eq{BFKL} over $b$ 
 the equation reduces
 to the BFKL equation\cite{BFKL} with the eigenfunctions
 $\Lb r^2\Rb^\gamma$. Therefore, the general solution takes the form:
 \beq \label{SOLBFKL}
 \int d^2 b\,N\Lb Y; \vec{x}_{01}, \vec{b}\Rb \,\,=\,\,\int^{\epsilon + i \infty}_{\epsilon - i \infty} \frac{ d \gamma}{ 2\,\pi\,i} e^{ \bas \chi\Lb \gamma\Rb\,\Lb Y - Y_0\Rb\,\,+\,\, \Lb \gamma\,-\,1 \Rb\xi}\, n_{in}\Lb \gamma\Rb
 \eeq
 where $\chi\Lb \gamma\Rb$ is given by \eq{KER} and
 $\xi\,=\,\ln\Lb 1/\Lb r^2\,Q^2_s\Lb Y_0\Rb\Rb\Rb$ with $r \equiv x_{01}$.
 In the definition of $\xi$ we introduce a new momentum scale which
 characterizes the value of the initial condition. For simplicity,
 we have that  
 \beq \label{DNAIC}
 N\Lb Y=Y_0,r\Rb \,=\,r^2 \,Q^2_s\Lb Y_0\Rb. 
 \eeq
 We can view this momentum as the saturation momentum at $Y=Y_0$, 
  since $N\, \sim \,1 $ at $r^2=1/Q^2_s(Y_0)$. \eq{DNAIC} leads
 to $n_{in} \,=\,\frac{1}{\gamma}$. 
 
 Finally,
  \beq \label{SOLBFKL1}
 \int d^2 b\,N\Lb Y; \vec{x}_{01}, \vec{b}\Rb \,\,=\,\,\int^{\epsilon + i \infty}_{\epsilon - i \infty} \frac{ d \gamma}{ 2\,\pi\,i} e^{ \bas \chi\Lb \gamma\Rb\,\Lb Y - Y_0\Rb\,\,+\,\, \,\Lb \gamma\,-\,1\Rb\xi}\, \frac{1}{\gamma }\,\,=\,\,\int^{\epsilon + i \infty}_{\epsilon - i \infty} \frac{ d \gamma}{ 2\,\pi\,i}\, e^{ \Psi\Lb Y,\xi,\gamma\Rb}\, \frac{1}{\gamma } 
  \eeq 
 
In the vicinity of the saturation scale, the integral over $ \gamma$ can
 be  evaluated using the method of steepest descent, with the 
equations for
 the saddle point $\gamma_{SP} \equiv  \gamma_{cr}$:
\beq \label{SP}
(1)~~~~~~\bas \chi\Lb \gamma_{cr}\Rb\,\Lb Y - Y_0\Rb\,\,+\,\, \Lb \gamma_{cr}\,-\,1\Rb\,\xi\,\,=\,\,0;~~~~
(2)~~~~\bas \frac{d \chi\Lb \gamma\Rb}{d \gamma}\Big{|}_{\gamma = \gamma_{cr}}\,\Lb Y - Y_0\Rb\,\,+\,\, \,\xi\,\,=\,\,0;
\eeq
Dividing the first equation by the second one, we obtain the value
 for $\bar{\gamma} = 1 - \gamma_{cr}$ 
which is the solution of \eq{GACR}. The first equation gives the
 value of 
the saturation momentum
\beq \label{EQQS}
\ln\Lb Q^2_s(Y)/Q^2(Y_0)\Rb\,=\,\bas \frac{\chi\Lb \gamma_{cr}\Rb}{1 - \gamma_{cr}} \Lb Y \,-\,Y_0\Rb \,\,=\,\,\lambda \,\Lb Y \,-\,Y_0\Rb
\eeq

Expanding $\Psi \Lb Y,\xi,\gamma\Rb$ at $\gamma \to \bar{\gamma}$ we obtain
\beq \label{DNAPSI}
\Psi \Lb Y,\xi,\gamma\Rb\,\,=\,\,\bar{\gamma} \,z \,\,-\Lb \gamma - \,\gamma_{cr}\Rb \,z\,\,+\,\,\h \bas \frac{d^2 \chi\Lb \gamma\Rb}{d \gamma^2}\Big{|}_{\gamma = \gamma_{cr}}\Lb \gamma - \gamma_{cr}\Rb^2\,\Lb Y - Y_0\Rb~~~\mbox{where}~~~z\,=\,
\ln \tau^2\,\,=\,\,\lambda\,\Lb Y - Y_0\Rb \,-\,\xi
\eeq

Plugging \eq{DNAPSI} into  \eq{SOLBFKL1}  and integrating
  over $\gamma -  \gamma_{cr}$, 
 the resulting solution can be written in the form
\beq \label{VQSMOD}
 N\Lb Y = \ln(1/x); r, b \Rb \,\,=\,\,\,N_0\, \Lb r^2\,Q_s\Lb Y, b\Rb\Rb^{\gamma_{\rm eff}}\,\,=\,\,N_0\, \Lb\tau\Rb^{
2\,\gamma_{\rm eff}} 
 \eeq
with
\beq \label{REPLACE}
\bar{\gamma}\,\,\longrightarrow\,\,\gamma_{\rm eff} \,\,=\,\,\bar{\gamma} \,\,+\,\,\frac{ \ln(1/\tau)}{\kappa\,\lambda \ln\Lb \frac{1}{x}\Rb }~~~~~~~\mbox{with}~~~~~\kappa\,=\,\frac{\chi''_{\gamma \gamma}\Lb \gamma\Rb}{\chi'_{\gamma }\Lb \gamma\Rb}\Big{|}_{\gamma=\gamma_{cr}}\,\approx \,9.9\,\, \mbox{in LO BFKL}
\eeq

In the method of  steepest descend, the saddle point for the integration
 turns out to be
\beq \label{SPZ}
\Lb \gamma - \gamma_{cr}\Rb_{SP}\,\,=\,\,\frac{z}{\bas \frac{d^2 \chi\Lb \gamma\Rb}{d \gamma^2}\Big{|}_{\gamma = \gamma_{cr}}\,\Lb Y - Y_0\Rb}
\eeq

\subsubsection{Impact-parameter dependent CGC dipole model}


As we have mentioned, the advantage of \eq{VQSMOD} is that we can
 introduce the correct behaviour
 of the amplitude at large impact parameter, by   imposing   the 
phenomenological
   decrease in  saturation momentum for large $b$, by writing it 
in the form:
\beq \label{QSB}
Q_s\,\,=\,\,Q_s\Lb x \Rb S\Lb b \Rb\,\,=\,\,Q_0 \Bigg(\frac{1}{x}\Bigg)^\lambda \,S\Lb b \Rb
\eeq
In the LO BFKL \cite{KOLEB}     $\lambda\,\, 
=\,\,\bas\, \frac{\chi\Lb \gamma_{cr}\Rb}{\bar{\gamma}}$.
Parameters $N_0$ and $Q_0$, as well as function $S\Lb b\Rb$
 should in future
 be taken  from non-perturbative QCD calculations but, at
 the moment, has to be determined from a fit to experimental 
 DIS data.
 We have two models\cite{RESH,CLP}\footnote{Actually, the same set
 of the data was described in the model of Ref.\cite{RSKV}, but this
 model does not include the correct behaviour  deep in the saturation
 region\cite{LETU},
 and we do not discuss it here.} on the market that describe
 the final
 set of the HERA experimental data on deep inelastic structure
 functions\cite{HERADATA} . They have different  forms for $S\Lb b 
\Rb$:
\bea \label{S}
\mbox{Ref\cite{RESH}}: & \to &   S\Lb b \Rb\,\,=\,\,\exp\Lb - \frac{b^2}{B}\Rb\,=\,\exp\Lb - \frac{b^2}{4\,\bar{\gamma}\,B_{\rm CGC}}\Rb;\label{RESHB}\\
\mbox{Ref\cite{CLP}}:& \to &   S\Lb b \Rb\,\,=\,\,\Lb m\,b\,K_1\Lb m\,b\Rb\Rb^{1/\bar{\gamma}};\label{CLPB}
\eea
The ansatz  of \eq{CLPB} is preferable, since it leads to
 $S\Lb b \Rb \xrightarrow{b \gg 1/m} \,\exp\Lb - m b\Rb$,
 which is  in accord with the Froissart theorem\cite{FROI} .
 However, we choose \eq{RESHB} which allows us to do several 
integrations analytically. Using \eq{I}, $\nabla^2_T N $ takes 
the following form, after integrating over $b$:

\bea \label{I}
\int d^2 b\,\nabla^2_T N \Lb Y, r, b  \Rb\,\,=\,\, \frac{1}{r^2} \left\{\begin{array}{l}\,8\, \pi \, B\, \bar{\gamma }\, N_0\, \tau ^{2 \bar{\gamma }} \Lb  1\,\,-\,\,N_0 \,\tau ^{2 \bar{\gamma }}\Rb; \,\,\,\mbox{for}\,\,\,\tau\,=\,r Q_s(x)\,<\,1\;\\ \\
 8\,\pi\,B\,\bar{\gamma}\,\phi_0 \tau^{2 \bar{\gamma}}\,\,\exp\Lb - 2\,\phi_0  \tau^{2 \bar{\gamma}}\Rb\,\,\,\,\, \mbox{for}\,\,\,\tau\,=\,r Q_s(x)\,>\,1;   \end{array}
\right.
\eea

with $\phi_0\,e^{ - 2 \phi_0}\,=\,N_0\,\Lb 1 - N_0\Rb$.

Before discussing the details of our model, we would like
 to outline,  which features of \eq{I} stem from the theorem,
 and which  from  phenomenological assumptions. The expression 
for $\tau \,\leq\,1$, as we have mentioned (see \eq{VQS})
 follows from the theory. However, the calculation of $N_G$
 (see \eq{NG}) takes into account the term of the order $N^2$.
 The corrections of the order of $N^2$ has to be estimated for
 $\tau \,<\,1$ and, in principle, they change  \eq{VSQ}.
 In Ref.\cite{GOLETE} we  discussed these corrections, but 
  we do  not take them into
 account, in \eq{I}  as we view $N = N_0 \tau^{2 \bar \gamma}$
 as a phenomenological expression that describes the DIS
 data\cite{RESH}. 

The fact that the impact parameter behaviour of the saturation
 momentum determines  the
$b$-dependence of the scattering amplitude,  comes from theory, while
 the particular form and result of  integration over $b$, stems
 from the model for $S\Lb b\Rb$. 

For $\tau \,\geq\,1$ we have discussed the form of \eq{I} in the 
 previous section (see also Ref.\cite{GOLETE}). The $b$ integration
 is  performed with the phenomenological
 $S\Lb b \Rb$.

In our estimates we use  the values of the parameters from 
Ref.\cite{RESH}(see Table 1). In this paper the HERA data were fitted
in the wide range of $Q^2$ from  $0.75\, GeV^2$ to $650 \, GeV^2$ .
 The expression for $Q_s(x) $ in this model is taken in the
 form\footnote{Note that we introduce the extra factor $\h$
 in the definition of the saturation scale,  since we use $\tau =
 r\,Q_s$, while in Ref.\cite{RESH} $\tau$ is defined as $\tau = r Q_s/2$.}
\beq \label{QSX}
Q_s\Lb x \Rb\,\,=\,\,\h \Lb \frac{x_0}{x}\Rb^{\frac{\lambda}{2}}\,GeV
\eeq

 It should be noted, that the value of $x$ from \eq{XF}  even at
  W = 13 TeV, is about $10^{-5}$,  which is in  the region that has been 
measured at HERA.
\begin{table}[h]
\begin{minipage}{13cm}{
\begin{tabular}{|l|l|l|l|l|}
\hline
$\bar{\gamma}$ & $N_0$  & $\lambda$ & $x_0$ & $B_{CGC}$ ($GeV^{-2}$)\\
\hline
 0.6599  $\pm$ 0.0003 & 0.3358$\pm$ 0.0004 & 0.2063 $\pm$ 0.0004 & $0.00105\pm 1.13 10^{-5}$ & 5.5  \\\hline
\end{tabular}
}
\end{minipage}
\begin{minipage}{4cm}
{\caption{Fitted parameters of the model\cite{RESH}, which we use for our
 estimates. }}
\end{minipage}
\label{t1}
\end{table}

\begin{boldmath}
\subsubsection{Inclusive production for $p_T \gg Q_s$}
\end{boldmath}


We can significantly simplify our calculation for $p_T \gg Q_s$. Indeed,
 for such large $p_T$ the integral over $r$ in \eq{MF} is concentrated
 in the region $\tau \ll 1$, where we can safely use the simple
 expression of \eq{DNA1}. The integral can be calculated explicitly
 and leads to the following
\bea \label{DNA3}
&&\frac{d \sigma_G}{d y \,d^2 p_{T}}\,\,=\\
&&\frac{2C_F}{\alpha_s 2\pi}\,\frac{1}{p^2_T} N_0^2 \,B^2 \,\bar{\gamma}^2\Bigg( \frac{\Gamma\Lb 2 \bar{\gamma} - 1\Rb}{\Gamma\Lb 2 - 2 \bar{\gamma}\Rb}\Lb \tilde{p}_T\Rb^{2\, -\, 4 \,\gamma_{\rm eff}}\,\,-\,\,N_0\,2^{ 2 \bar{\gamma} +1}\frac{\Gamma\Lb 3 \bar{\gamma} -1\Rb}{\Gamma\Lb 2 - 3 \bar{\gamma}\Rb}\Lb \tilde{p}_T\Rb^{2 \,-\, 6 \,\gamma_{\rm eff}} \,\,+\,\,N^2_0\,2^{2 \bar{\gamma}}\frac{\Gamma\Lb 4 \bar{\gamma} - 1\Rb}{\Gamma\Lb 2 - 4 \bar{\gamma}\Rb}\Lb \tilde{p}_T\Rb^{2\, -\, 8 \,\gamma_{\rm eff}}\Bigg)\nn
\eea

\begin{figure}[ht]
   \centering
  \leavevmode
   \includegraphics[width=7.5cm]{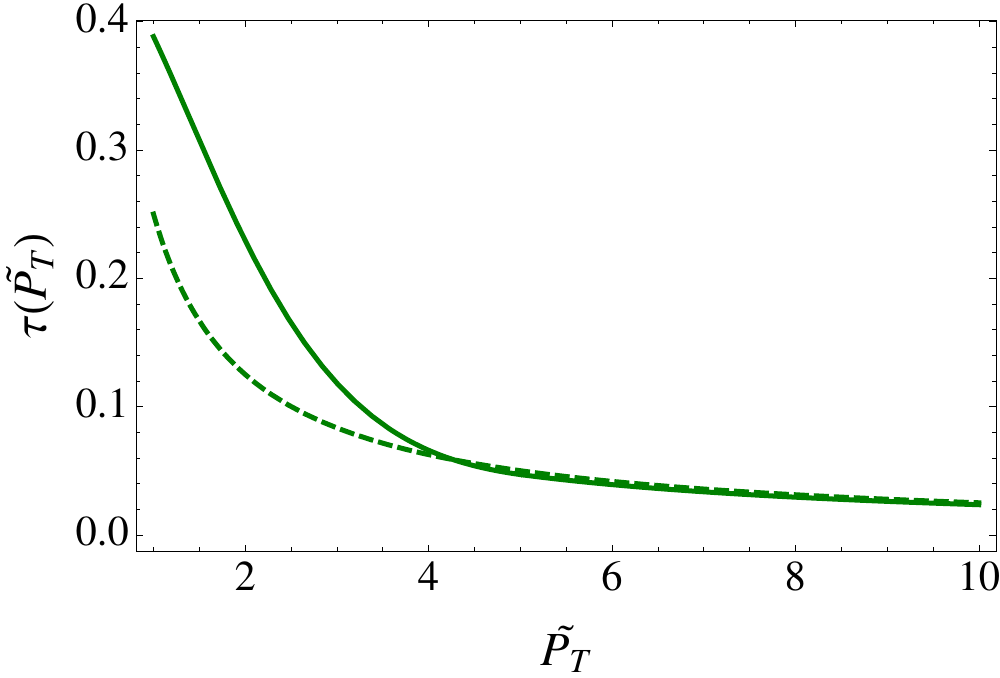} \protect\caption
{ The average $\tau$ (see \eq{AVT}). The solid line is the
 result of  numerical calculations. The dotted line shows that at large
 $\tilde{p}_T$  \,\,$\langle \tau \rangle \, \propto\, 1/(4 \tilde{p}_T)$.
 The figure is taken from Ref.\cite{GOLETE}.    $\tilde{p}_T \,=\,p_T/Q_s$}
\label{tau}
   \end{figure}

 We know that in the vicinity of the saturation scale
 the scattering amplitude in the momentum representation 
has the following behaviour:
\beq \label{VQSM}
N\Lb p_T\Rb\,\,=\,\,{\rm Const} \Lb \frac{p^2_T}{Q^2_s(x)}\Rb^{\bar{\gamma}}
\eeq
Therefore, from  \eq{DNA3} we can determine  the value of the constant in 
\eq{VQSM} from the value of $N_0$.

\eq{DNA3} allows us to take into account the violation of the geometric
 scaling behaviour ,  given by 
\eq{VQSMOD} and \eq{REPLACE}. We can make  such a replacement  directly
 in the momentum
 representation, since $r  \propto 1/p_T$. However, we need to find
 the coefficient in front of $p_T$ and, perhaps, an additional
 constant. We calculate
the average $\tau$  using the expression:
\beq \label{AVT}
\langle \tau \rangle \Lb \tilde{p}_T\Rb\,=\,\frac{ \int
 \tau J_0\Lb \tilde{p}_T \tau\Rb I\Lb \tau\Rb d \tau}{ \int  J_0\Lb \tilde{p}_T \tau\Rb I\Lb \tau\Rb d \tau}~~~~~~~~
 \mbox{where}~~~~~~I\Lb \tau\Rb\,\,=\,\, \int d^2 b\, S\Lb b \Rb \Bigg( \frac{d^2 N_G\Lb z(r, b, x_1))\Rb}{d z^2}\Bigg)^2\eeq
The results of these estimates are shown in \fig{tau}.
 One can see at at 
$\tilde{p}_T \to 0$  
$\langle \tau \rangle\,=\,0.478 \approx \h$ while at
 large $\tilde{p}_T$ it is proportional to $1/(4 \tilde{p}_T)$.

  For $\tilde{p}_T=0$, or more generally for $p_T \ll Q_s$,  the
 typical distance turns out to be $r =1/(2 Q_s(x))$, and for large
 $\tilde{p}_T$ {\bf it is} of the order of $1/(4 p_T)$.
Hence, we suggest to use in \eq{REPLACE}   the calculated $\langle
 \tau \rangle\Lb \tilde{p}_T\Rb$ for $\tilde{p} \leq 4$ and $1/(4
 \tilde{p}_T$) for $\tau \geq 4$ and to  substitute in \eq{DNA3} :
 \beq \label{TR}
\gamma_{\rm eff}\Lb \tilde{p}_T, x\Rb \,\,=\,\,\bar{\gamma} \,\,+\,\,\frac{ \ln(1/\langle \tau \rangle \Lb \tilde{p}_T\Rb)}{\kappa\,\lambda \ln\Lb \frac{1}{x}\Rb } 
\eeq
 where $\tilde{p}_T\,=\,p_T/Q_s$.

In Re.\cite{GOLETE} it was shown that this replacement results in
 the $\gamma_{\rm eff} \,>\, 1 $    at large $\tilde{p}_T$. This
 dependence is essential for
 describing the experimental data. Indeed, the value of $n$
 in the hard term in \eq{SUM} is $n=3.1$. As we  have  seen above
 (see \eq{DNA3} for example) at large $p_T$ the inclusive cross
 section is proportional to $1/p_T^{4 \gamma_{\rm eff}}$. For
 $\gamma_{\rm eff} = \bar{\gamma} $ it is impossible to obtain
 a decrease of about $1/p^6_T$, as  indicated by the data, while \eq{TR}
 makes such a description possible
  (see Ref.\cite{GOLETE} for detail discussion).

\begin{boldmath}
\subsection{Dipole-nucleus scattering amplitude}
\end{boldmath}

  
  For  the dipole-nucleus scattering amplitude in the vicinity of the
 saturation scale $\tau \,\sim \,1$ we have the same behaviour as
 for the dipole-nucleon amplitude:
  \beq \label{DAAVQS}
  N^{A}\Lb Y,r, b\Rb\,\,=\,\,{\rm Const} \Lb r^2\,Q_S\Lb A,  Y, b\Rb\Rb^{\bar{\gamma}}
  \eeq
  The main question that we will try to answer in this section is
 how the value of ${\rm Const}$ in \eq{DAAVQS} is related to $N_0$
 for the dipole-nucleon scattering amplitude.
  
\begin{figure}[ht]
   \centering
  \leavevmode   
  \begin{tabular}{c c c}  
      \includegraphics[width=8cm]{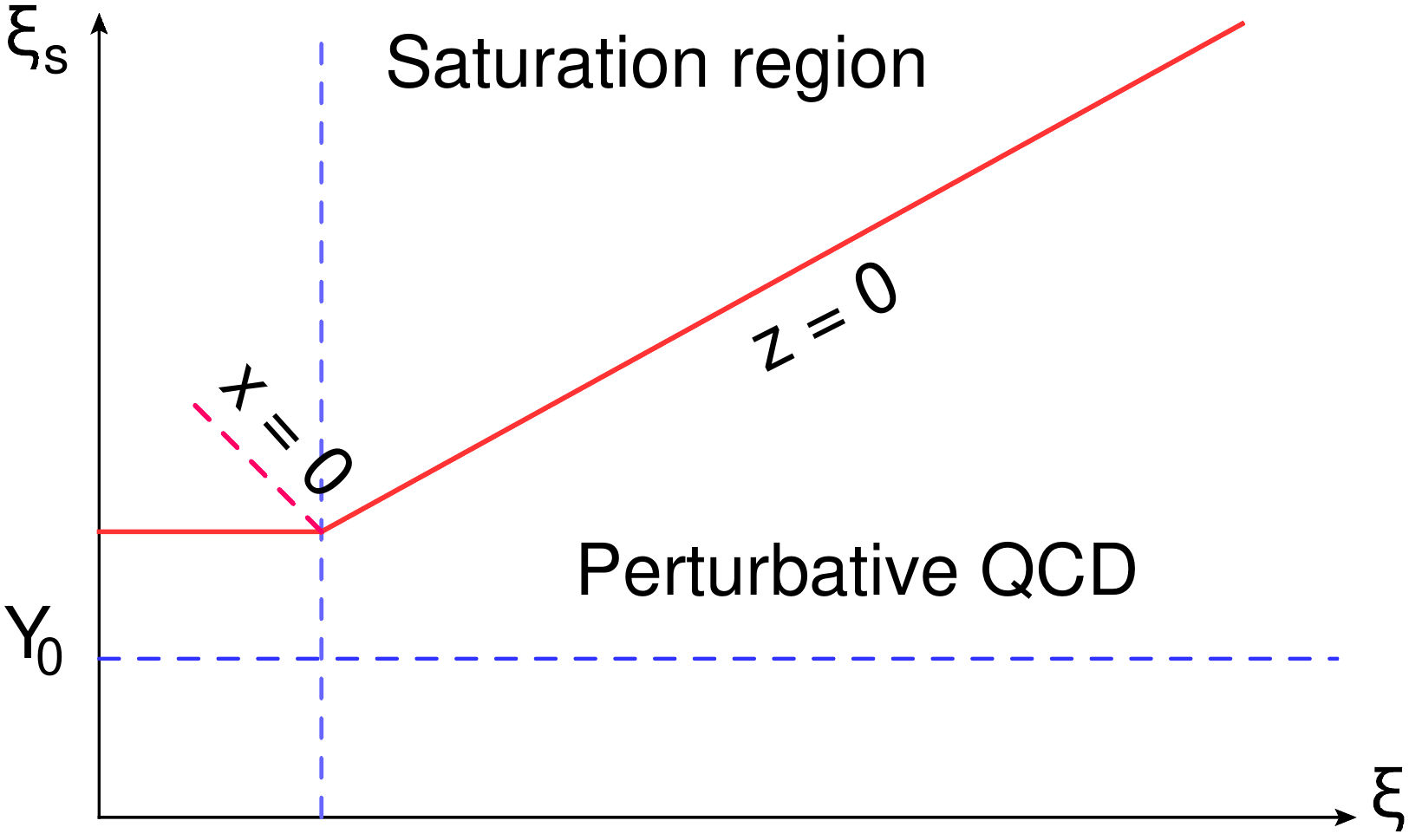}  &~~~~~~~ ~~~~~~~&
        \includegraphics[width=8cm]{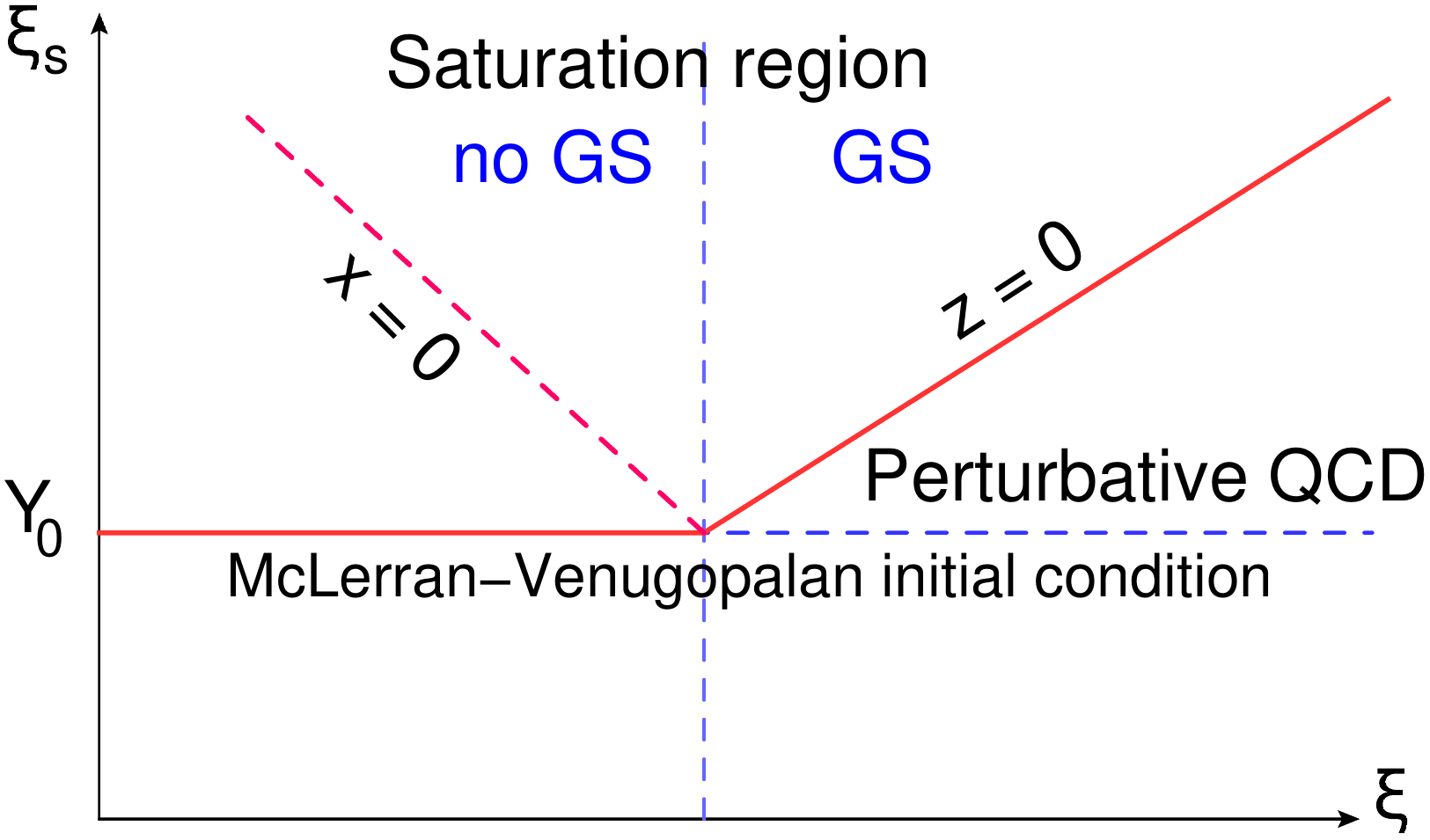} \\
      \fig{map}-a & &    \fig{map}-a\\
      \end{tabular}
             \caption{The QCD map. \fig{map}-a shows the kinematic regions
for the dipole-nucleon amplitude, while
 in \fig{map}-b, we show  the kinematic regions for  dipole-nucleus 
scattering.
  Note  that the saturation domain in this
 case can be divided in two subregions: (i) for $\xi < 0$,  where
 we expect  geometric scaling behaviour of the scattering
 amplitude, and $\xi > 0$ where there is no such
 behaviour. $\xi_s \,=\,\lambda\,\Lb Y - Y_{\rm min}\Rb$ and $\xi\,=\,-\ln\Lb \tau^2\Rb\,=\,\ln\Lb r^2\,Q^2_s\Rb$. $z = \xi_s\, -\, \xi$ and $ x \,=\,\xi\,+\,\xi$.}
\label{map}
 \end{figure}

  The principle difference between scattering with a nucleus and a nucleon
 is shown in \fig{map}. For  dipole-nucleon scattering the initial
 condition at $Y\,=\,Y_0 \equiv Y_{\rm min}$ is such that
 $N\Lb Y=Y_0,r, b\Rb \,\ll\,1$(see \fig{map}-a). On the other hand,
 for dipole-nucleus scattering even at $Y=Y_0$ the shadowing 
corrections are large, and  we impose the initial conditions
 inside the saturation region (see \fig{map}-b). These  initial
 conditions lead to different solutions for dipole-hadron and
 dipole-nucleus amplitude in the saturation region (see
 Refs.\cite{GOLEODD,CLM,KLTA,LETU1}). In particular, for $\xi < 0$,  where
 we expect  geometric scaling behaviour of the scattering
 amplitude, and for $\xi > 0$  and no such behaviour is seen.  
 However, a glance at \fig{map}, shows that for
 $Y - Y_{\rm min} \,\gg\,1$ for $\nabla_T^2 N^{A}\Lb Y,r, b\Rb  
 $ we do not expect  violation of the geometric
 scaling behaviour of the scattering amplitude, since,
 as has been discussed in the previous section, this
 observable gives the main contribution in the vicinity of
 the saturation scale shown by 
red line in \fig{map}. As we have 
discussed for the case of the dipole-nucleon scattering, the behaviour
 of the scattering amplitude in the vicinity of the saturation scale
 is determined by the solution to the linear BFKL equation (see \eq{BFKL}).


Therefore, to find the value of ${\rm Const} $ in \eq{DAAVQS}, we need to
 solve the BFKL equation with the correct initial condition. This 
condition is given by the McLerran-Venugopalan formula \cite{MV}, which we 
use in the simplified form;
\beq \label{MV}
N^A_{\rm in}\Lb Y=Y_0; r, b\Rb\,\,=\,\,1 \,\,-\,\,\exp\Bigg(- r^2 \,Q^2_s\Lb A, Y=Y_0,b \Rb\Bigg)
\eeq

The general solution to the BFKL equation of \eq{BFKL} has the same
 form as in \eq{SOLBFKL}:
\beq \label{NAASOL} 
N^A\Lb Y; \vec{x}_{01}, \vec{b}\Rb \,\,=\,\,\int^{\epsilon + i \infty}_{\epsilon - i \infty} \frac{ d \gamma}{ 2\,\pi\,i} e^{ \bas \chi\Lb \gamma\Rb\,\Lb Y - Y_0\Rb\,\,+\,\, \Lb \gamma\,-\,1 \Rb\xi}\, n^A_{in}\Lb \gamma, \frac{Q^2_s\Lb A, Y_0,b\Rb}{Q^2_s\Lb  Y_0\Rb}\Rb \eeq
where $Q_s\Lb Y_0\Rb$ has been introduced in \eq{DNAIC}. We recall
 that $\xi \,=\,-\,\ln\Lb r^2 Q^2_s\Lb Y_0\Rb\Rb$. Using \eq{MV} we
 can find $n^A_{in}\Lb \gamma, \frac{Q^2_s\Lb A, Y_0,b\Rb}{Q^2_s\Lb  Y_0\Rb}\Rb$ we takes the following form:

\beq \label{NIN}
n^A_{in}\Lb \gamma, \frac{Q^2_s\Lb A, Y_0,b\Rb}{Q^2_s\Lb  Y_0\Rb}\Rb\,\,=\,\,\Lb  \frac{Q^2_s\Lb A, Y_0,b\Rb}{Q^2_s\Lb  Y_0\Rb}\Rb^{1\,-\,\gamma} \,\Bigg( \Gamma\Lb \gamma \Rb\,\,-\,\,\Gamma\Lb \gamma, \frac{Q^2_s\Lb A, Y_0,b\Rb}{Q^2_s\Lb  Y_0\Rb}\Rb\Bigg)
\eeq

Plugging \eq{NIN} into \eq{NAASOL} we obtain the following solution:
\beq \label{NAASOL1} 
N^A\Lb Y; \vec{x}_{01}, \vec{b}\Rb \,\,=\,\,\int^{\epsilon + i \infty}_{\epsilon - i \infty} \frac{ d \gamma}{ 2\,\pi\,i} e^{ \bas \chi\Lb \gamma\Rb\,\Lb Y - Y_0\Rb\,\,+\,\, \Lb \gamma\,-\,1 \Rb\xi_A}\, \Bigg( \Gamma\Lb \gamma \Rb\,\,-\,\,\Gamma\Lb \gamma, \frac{Q^2_s\Lb A, Y_0,b\Rb}{Q^2_s\Lb  Y_0\Rb}\Rb\Bigg)
 \eeq
where $\xi_A\,\,=\,\,-\,\ln\Big( r^2 Q^2_s\Lb A, Y_0, b\Rb\Big) $.

Solving \eq{SP}, we obtain the solution in the vicinity of saturation,
 in the form of \eq{DAAVQS} scale with
\beq \label{CONST}
{\rm Const}\,\,=\,N_0 \frac{\Bigg( \Gamma\Lb \gamma_{cr} 
 \Rb\,\,-\,\,\Gamma\Lb \gamma_{cr}, \frac{Q^2_s\Lb A, Y_0,b\Rb}{Q^2_s\Lb
  Y_0\Rb}\Rb\Bigg)}{\Bigg( \Gamma\Lb \gamma_{cr}  \Rb\,\,-\,\,\Gamma\Lb
 \gamma_{cr}, 1\Rb\Bigg)}\,\,=\,\,N_0\,R
\eeq

\begin{figure}[ht]
   \centering
  \leavevmode   
  
  \begin{tabular}{c c c}  
      \includegraphics[width=8cm]{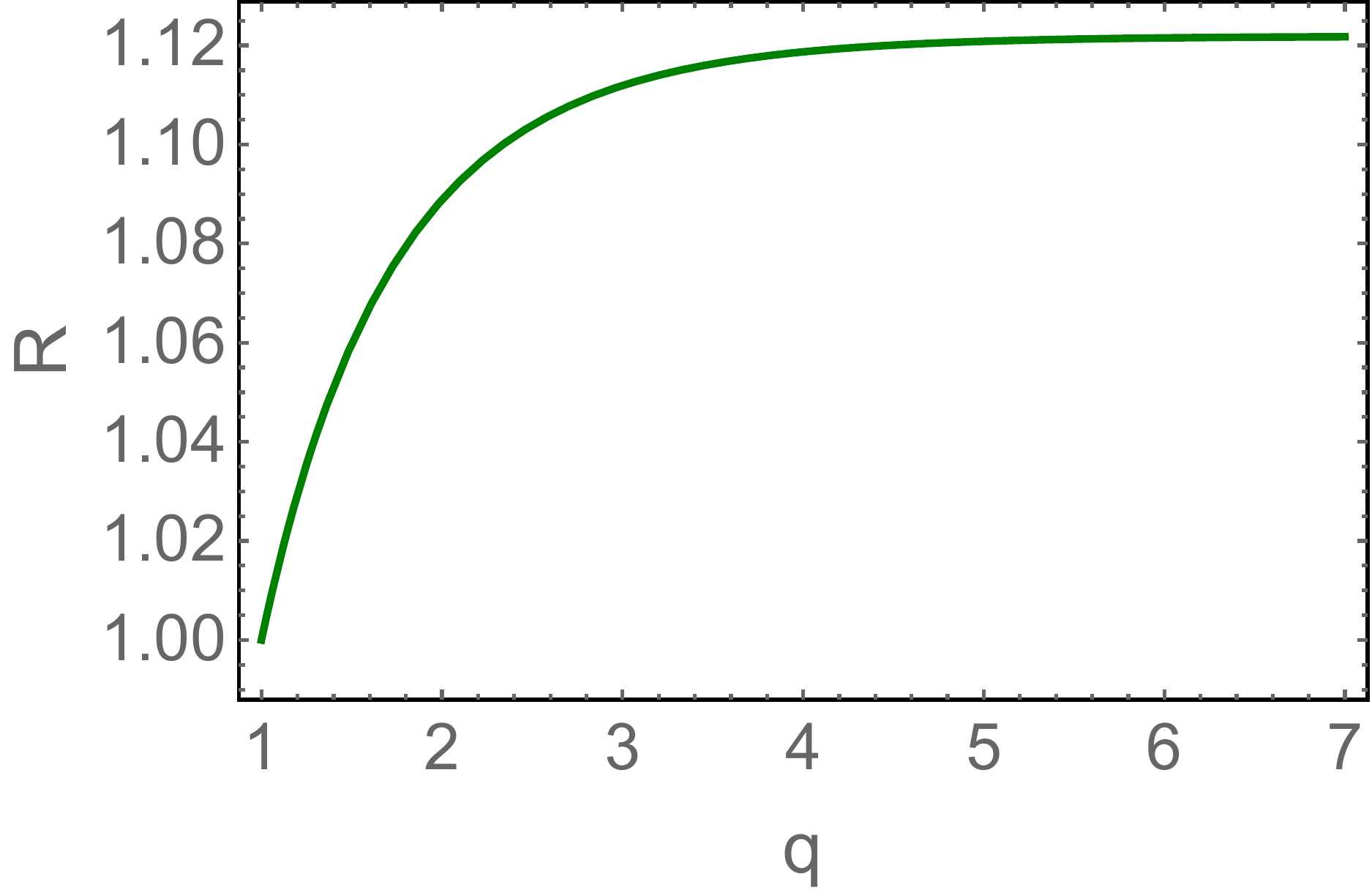}  & ~~~~    &\includegraphics[width=8cm]{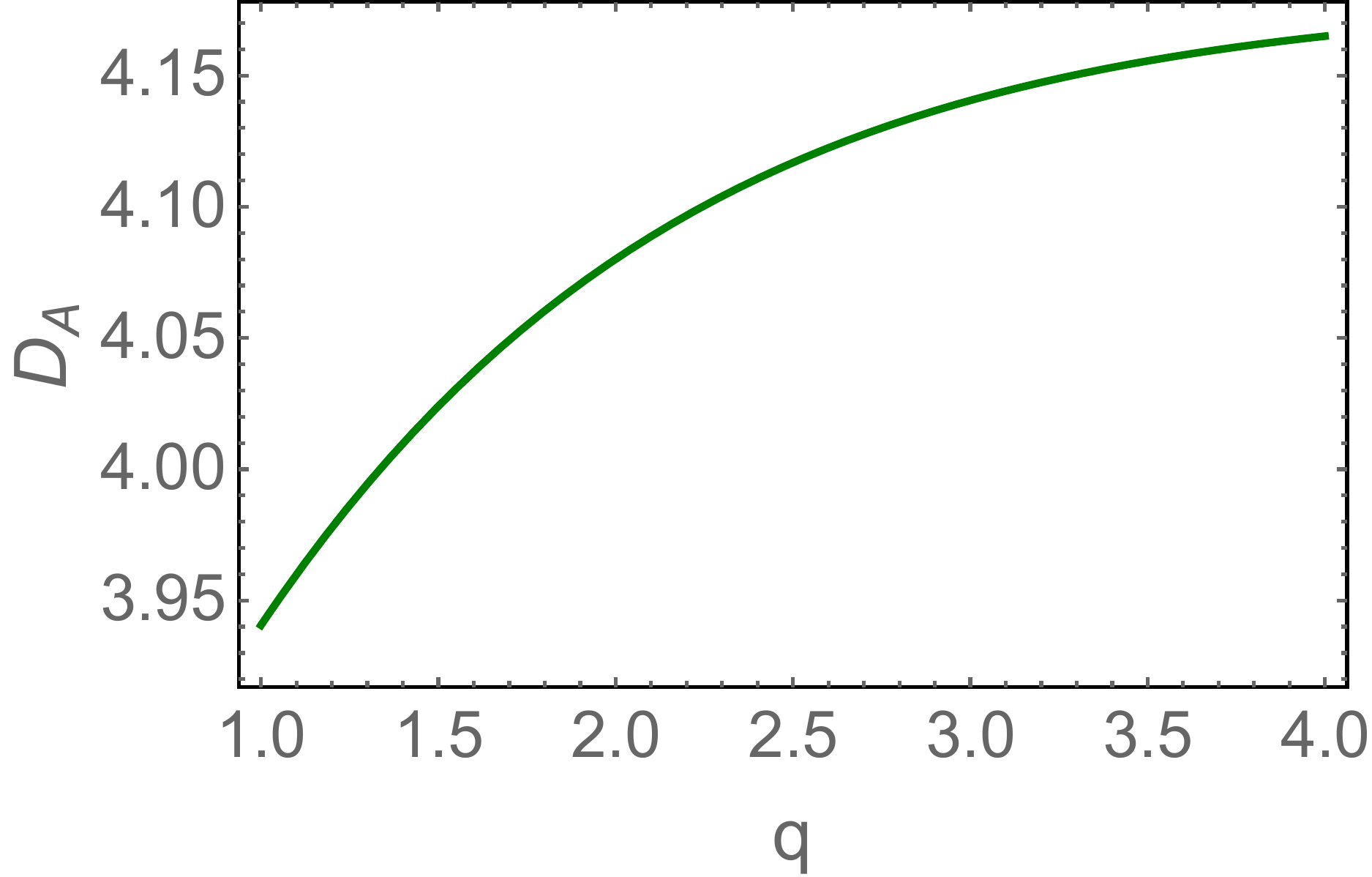}\\
      \fig{q}-a & & \fig{q}-b\\
      \end{tabular}
        \caption{\fig{q}-a:  R versus $q = Q^2_s\Lb A, Y_0, b\Rb{\Big{/}}
Q^2_s\Lb  Y_0\Rb$ (see \eq{CONST}).\fig{q}-b: $D_A$ as function of $q$.}
\label{q}
 \end{figure}

In \fig{q} we plot the value of R as function of
 $q = Q^2_s\Lb A, Y_0, b\Rb{\Big{/}}Q^2_s\Lb  Y_0\Rb$. One can see
 that $R$ is very close to 1.  The chosen range of $q$ we will be
 discussed in the next section. 
 In further estimates,  we take ${\rm Const} = N_0$.

As we  discussed in the previous section, \eq{TR} which takes into
 account the violation of the geometric scaling behaviour, is
  essential  for the  description of the experimental data, since the
 region of perturbative QCD gives a large contribution.  Bearing this
 in mind, we take into account the contribution of $n^A_{in}\Lb \gamma,
 \frac{Q^2_s\Lb A, Y_0,b\Rb}{Q^2_s\Lb  Y_0\Rb}\Rb$ into effective
 $\gamma_{\rm eff}$ as well as \eq{TR}. Considering the resulting
 $\Psi$ in \eq{SOLBFKL1} in the form
\beq \label{PSIA}
\Psi\Lb Y, \xi_A, \gamma\Rb\,\,=\,\, \bas \chi\Lb \gamma\Rb\,\Lb Y - Y_0\Rb\,\,+\,\, \,\Lb \gamma\,-\,1\Rb\xi\,+\,\ln\Lb n^A_{in}\Lb \gamma, \frac{Q^2_s\Lb A, Y_0,b\Rb}{Q^2_s\Lb  Y_0\Rb}\Rb\Rb
\eeq
we obtain  the following expression for the effective $\gamma$: 

\beq \label{GAEFFA}
\gamma_{\rm eff} \,\,=\,\,\bar{\gamma} \,\,+\,\,\frac{ \ln(1/\tau)}{\kappa\,\lambda \ln\Lb \frac{1}{x}\Rb }\,\,+\,\,2\,D_A \frac{ \ln(1/\tau)}{\Lb \kappa\,\lambda \ln\Lb \frac{1}{x}\Rb\Rb^2 }
\eeq
where $\kappa$  is defined in \eq{TR}. Recall that $\tau = r \,Q_s\Lb A,Y\Rb$.
 The value of $D_A$ is plotted in \fig{q}-b,
as a function of $q = Q^2_s\Lb A, Y_0, b\Rb{\Big{/}}Q^2_s\Lb  Y_0\Rb$.
In the region of $q = 1.5 - 4 $ $D_A \approx 4$, and we will argue
 below that  this region contributes to the ion-ion collisions
 at the LHC energies.

~~

\section{KLN model for ion-ion scattering at high energy}

\begin{figure}[ht]
\begin{center}
 \includegraphics[width=9cm]{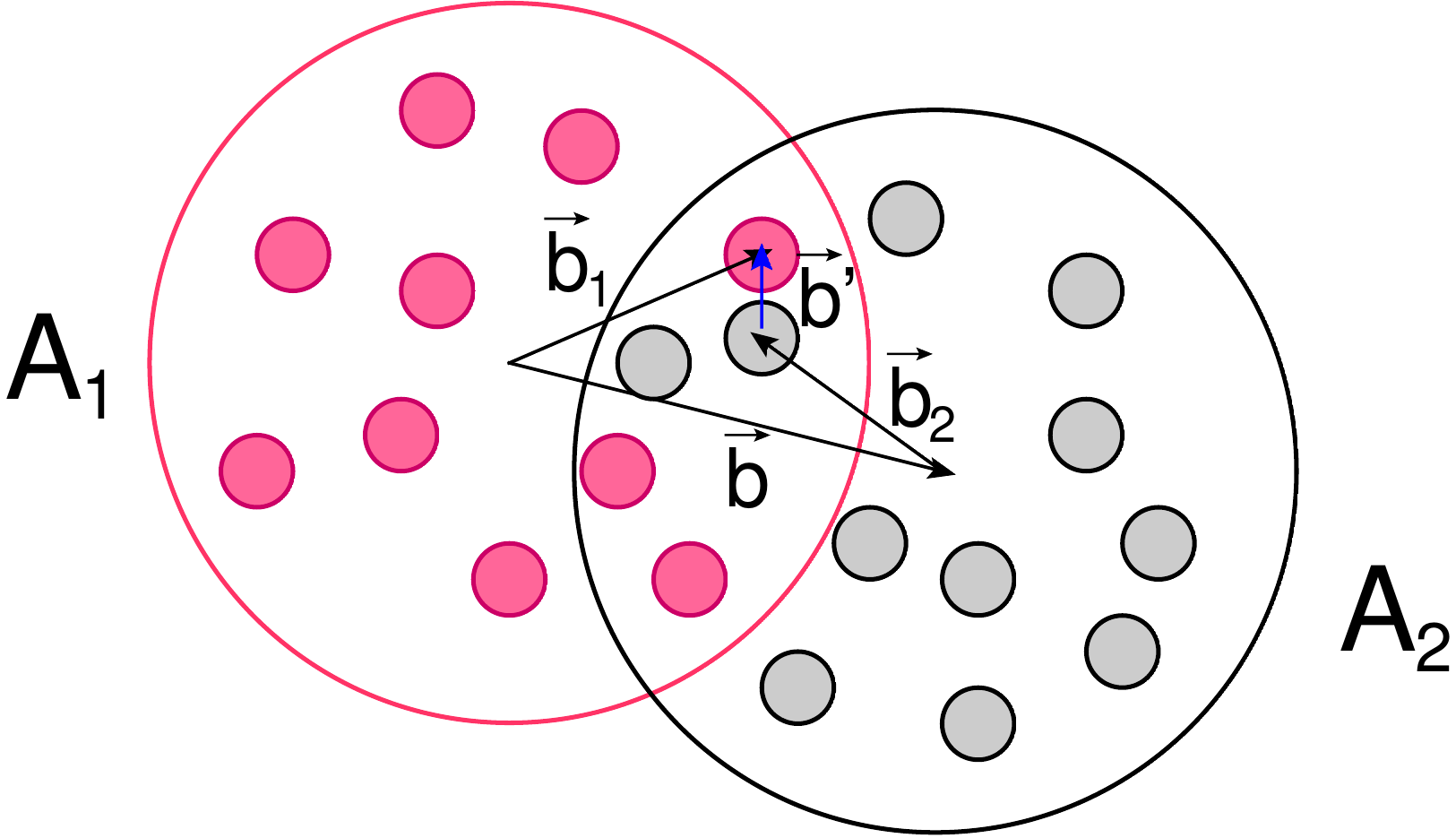}
\end{center}
    \protect\caption{ The kinematics for the nucleus-nucleus interaction.
 $\vec{b}_1 $ denotes the position of the nucleon in the transverse plane
 in the nucleus $A_1$. 
$\vec{b}_2$  the position of the   nucleon in  nucleus $A_2$. $b$ 
denotes the distance
 between the centers of two nuclei
 $\vec{b} \,=\,\vec{b}_1\,-\,\vec{b}_2$.	
 The two nucleons from different nuclei, which interact at distance   
$\vec{b}'$, 
 are shown by different colours
 in the figure.
}
\label{aakin}
   \end{figure}

\fig{aakin} shows the interaction of two nucleus $A_1$ and $A_2$. From
 this picture we see that we need to know how many nucleons interact,
 and how this interaction occurs. In particular, one nucleon can
 interact with several nucleons from another nucleus, and so on. We
  use the KLN model to answer all similar questions (see Refs.
\cite{KLN1,KLN2,KLN3,KLN4,KLN5,DKLN}).  We choose this model since, on
 one hand, this model not only successfully described the RHIC data  
and
 it's predictions for  the experimental data at the LHC, are quite good. 
On the other
 hand, it is  based on the Glauber approach\cite{GLAUB,MRSS}, which
 successfully describes the nucleus-nucleus interaction over a  wide
  energy range. In Refs.\cite{KLNS,KLN1}  the Glauber approach is
 developed for the hadron production for nucleus-nucleus collisions, 
 and it  demonstrates how we can  calculate (i)  the number of
 nucleons that participate in the interactions (number of participants
 $N_{\rm part}$), their density $\rho_{\rm part}$ at fixed values of
 $b$; and (2) how to correlate the typical values of $b$ with the
 centrality $c$,  which is measured experimentally, and which
 characterizes the percentile of events with the largest number
 of produced particles (as registered in detectors).
 Experimentally collisions are grouped into event(centrality) classes,
 with the most central class defined by events with the highest
 multiplicity(smallest forward energy), which corresponds to small
 values of the impact parameter.  Basically centrality $c\Lb N\Rb$
 is equal to
 \beq \label{C}
 c\Lb N\Rb\,\simeq\,\,\frac{\pi b \Lb N\Rb}{\sigma_{n}}\eeq, 
 where $c\Lb N\Rb$ is the centrality of the events with the multiplicity
  higher than $N$ .$b\Lb N\Rb$ is the value of the impact parameter for
 which the average multiplicity $n\Big(b\Lb N\Rb\Big) = N$.

In our approach,  we use the estimates of Ref.\cite{KLN1} for the density
 of the participants at for different centrality classes\footnote{For
 comparison with the ALICE experimental data we use the results of the
 estimates, given by Ref.\cite{ALICEC}, in which the procedure of
 Ref.\cite{KLN1} is used in the Glauber Monte Carlo.}.  

The key idea of the KLN model \cite{KLN1,KLN2}  is,  that the
 saturation momentum is proportional to  the density of the
 participants, $Q_s \,\propto\,\,\rho_{\rm part}$. The arguments
 for this stem from the simple equation for the saturation
 momentum\cite{KOLEB,GLR,MUQI,MV}

\beq \label{QSKLN}
Q^2_s\Lb A,Y\Rb\,\,=\,\,\frac{ 8 \,\pi^2\,N_C}{N^2_c - 1} \,\as\Lb Q_s^2\Rb \frac{x \,G_A\Lb Q^2_s\Lb A, Y\Rb, Y\Rb}{\pi\,R^2_A}
\eeq
where $Y \,=\,\ln\Lb 1/x\Rb$, $N_c $ denotes the number of colours, $R_A$ 
 the radius of the nucleus and $xG_A$ is the gluon structure function of
 the nucleus.

For the kinematic region where the gluon densities are small we can safely 
consider $G_A\Lb Q^2, x\Rb\,\,=\,\,A\,xG_N \Lb Q^2,x\Rb$, where $G_N$ is
 the gluon structure function for a nucleon. Plugging  this 
relation in \eq{QSKLN} 
  as well as $R_A \,=\,R_N A^{1/3}$, where $R_N$ is the radius
 of the nucleon, we obtain that
\beq \label{QSKLN1}
Q^2_s\Lb A,Y\Rb\,\,=\,\,\frac{ 8 \,\pi^2\,N_C}{N^2_c - 1} \,\as\Lb Q_s^2\Rb \frac{A}{\pi R_A} x \,G_N\Lb Q^2_s\Lb A, Y\Rb, Y\Rb\,\,=\,\rho_A \,\frac{ 8 \,\pi^2\,N_C}{N^2_c - 1} \,\as\Lb Q_s^2\Rb  x \,G_N\Lb Q^2_s\Lb A, Y\Rb, Y\Rb\eeq
where $\rho_A$ is the density of the nucleons in the
 nucleus in the transverse plane.

The KLN model generalizes \eq{QSKLN1} proposing
\beq \label{QSKLN2}
Q^2_s\Lb N_{\rm part},Y\Rb\,\,=\,\,\h \rho_{\rm part}\,\,\frac{ 8 \,\pi^2\,N_C}{N^2_c - 1} \,\as\Lb Q_s^2\Rb  x \,G_N\Lb Q^2_s\Lb N_{\rm part}, Y\Rb, Y\Rb
\eeq
We suggest to re-write \eq{QSKLN2} in the form
\beq \label{QSKLN3}
Q^2_s\Lb N_{\rm part},Y\Rb\,\,=\,\,\h \rho_{\rm part}\,\,\pi R^2_N\,\fbox{\large$\frac{ 8 \,\pi^2\,N_C}{N^2_c - 1} \,\as\Lb Q_s^2\Rb   \,\frac{x\,G_N\Lb Q^2_s\Lb N_{\rm part}, Y\Rb, Y\Rb}{\pi\,R^2_N}$}\,\,=\,\,\h \rho_{\rm part}\,\int d^2 b \,\,Q^2_s\Lb N, Y, b\Rb
\eeq

Factor $\h$ reflects  the fact that we are dealing with the density  
of those
 nucleons in a single
nucleus, which will participate in the collision at a given impact
 parameter $b$, or in a definite centrality class. In \eq{QSKLN3} we
 write the expression for the nucleon saturation momentum in the 
frame.

One can see that for the DIS with a nucleus with $N_{\rm part} = A$ we obtain 
from \eq{QSKLN3} 
\beq \label{QSKLNA}
Q^2_s\Lb A,Y\Rb\,\,=\,\,A^{1/3} \,Q^2_s\Lb N, Y\Rb
\eeq
in accord with Refs.\cite{MU03,KLM,MU99,LERYA}. Using the estimstes
 of $\rho_{\rm part}$ from Ref.\cite{KLN1} we obtain that the range
 of $q$ in \fig{q} for the lead-lead scattering  is $q = 1 - 3.06$.

 For the gluon transverse momenta $p_T$  
 distributions,  we suggest
 the following formula for the ion-ion collisions in the definite
 centrality class with the number of participants $N_{\rm part}$:
\bea \label{MFA}
&&\frac{d n_G}{d y \,d^2 p_{T}}\Big{|}_{N_{\rm part}}\,\,=\h \,N_{\rm part} \frac{d N}{d y d^2 p_T}\Big{|}_{\rm proton-proton}\,\,=\,\,\h \,N_{\rm part} \,\frac{1}{\sigma_{in}}\,\frac{d \sigma^N_G}{d y \,d^2 p_{T}}\,\,\\
&& = \h \,N_{\rm part} \,\frac{1}{\sigma_{in}}\,\,\frac{2C_F}{\alpha_s (2\pi)^4}\,\frac{1}{p^2_T} \,\int d^2  r\,e^{i \vec{p}_T\cdot \vec{r}}\,\,\,\int d^2 b\,\nabla^2_T N_G\Lb y_1 = \ln(1/x_1); r, b \Rb\,\,\int d^2 b' \,\nabla^2_T\,N_G\Lb y_2 = \ln(1/x_2); r,  b' \Rb. \nn
\eea
where  $N_G$ is related to the scattering dipole-nucleon amplitude
 (see \eq{NG}), and $n_G$ denotes the multiplicity of the emitted gluons.

$\sigma_{in}$ denotes  the inelastic cross section for the nucleon-nucleon
 scattering at  corresponding energy.  One can see, that in \eq{MFA}
 we consider  each participant as a nucleon which interacts with
 another nucleon from the different nucleus, at  impact parameter
 $b'$ (see \fig{aakin}).
  The saturation momentum for this scattering  is given by
 \eq{QSKLN3},  for the effective $\gamma_{\rm eff}$ we take 
\eq{GAEFFA}, which includes corrections from the interaction
 with the other nucleons. We need also to fix the dependence
 of the saturation scale on the impact parameter of the
 nucleon-nucleon interaction ($b'$ in \fig{aakin}). Finally,
 the equation for the saturation scale, which we use in our
 estimates of gluon production in nucleon-nucleon scattering,  takes the form:
 
 \beq \label{QSAA}
 Q^2_s\Lb N_{\rm part}, Y, b'\Rb\,\,=\,\,\h \rho_{\rm part}\,\int\,d^2 b\,Q^2_s\Lb Y, b  \Rb\,S\Lb b'\Rb
 \eeq
 where $Q_s$ and  $S\Lb b'\Rb $ are  given by \eq{QSB} and \eq{RESHB},
 respectively.

At first sight, we do not need \eq{MFA}, since we can use the master
 equation ( see \eq{MF}) and using the  experimental information of
 gluon structure functions of nucleus (see for example Ref.\cite{AKER}),
  we can calculate the inclusive cross section as we did for proton-proton
 scattering in Ref.\cite{GOLETE}.  Indeed, for the inclusive production of
 gluons for an ion-ion collision, we can proceed in this manner, but the
 factorization of \eq{MF} is proven only in the case, when we did not
 make any additional selections of the events summing over all
 accompanying hadrons. Considering the different centrality
 classes we make the additional selection on the multiplicity
 of produced hadrons. It is sufficient to refer to the AGK cutting
 rules\cite{AGK}, to see that these selections violate the
 factorization of \eq{MF}.  The violation of the AGK cutting
 rules in QCD (see Ref.\cite{KOLEB}) makes the situation even
 worse. Hence, the KLN model gives us the simple way to estimate
 the cross sections for the different centrality classes.

\section{Comparison with the experiment}


We calculate the cross section for gluon production using \eq{MFA}.
 As we have discussed in section II-B, the typical values of $r$ 
 that contribute to the integral in \eq{MFA} are rather small, of
 the order of $Q_s$. This means that we can safely use the CGC/saturation
 approach or/and the perturbative QCD estimates for this integral. We do 
not need to incorporate
 modifications of the gluon propagators, for example due to the
 confinement \cite{VAZW,KHALE},  in
 our calculations.

   The factor $1/p^2_T$ in front  in \eq{MFA} stems
 from the gluon propagator \cite{BFKL}, and it is affected both by the
 hadronization, and by  interactions with co-movers in the parton
 cascade.  In our approach to the confinement problem, we first 
  need to take into account,   the effect of the mass of produced
 gluon jet due to hadronization, which changes the gluon
 propagator\cite{KLN2}:
\beq \label{GLPROP}
G\Lb p_T\Rb\,\, =\,\, \frac{1}{p^2_T}\,\,\longrightarrow\,\,\frac{1}{p^2_T \,\,+\,\,2 \Big( Q_s \Theta\Lb Q_s - p_T\Rb\,+\,p_T \Theta\Lb p_T - Q_s\Rb\Big)\,m_{\rm eff}}
\eeq
Therefore, we calculate  gluon production using \eq{MFA}, in which
 we use \eq{GLPROP} to replace the factor $1/p^2_T$. Therefore, our
 model of the hadronization consists of two ingredients: the decay
 of the gluon into hadron jet with the fragmentation function (see
 \eq{XSPION}), and  to  account for the mass of this jet, using 
\eq{GLPROP}.
 It is instructive to note, that in our approach we see two stages of the
 multi-particle production in an explicit way: the creation of the
 Colour Glass Condensate (calculation of the integral over $r$ in
 \eq{MFA}) and the stage of hadronization ( \eq{GLPROP}, and
 fragmentation functions).

For calculating the integral over $r$ in \eq{MFA}  we use
 \eq{I} for $p_T \,\leq\,2\,Q_s(x)$ and \eq{DNA3} for
 larger values of $p_T$ (see Ref.\cite{GOLETE} for more details).  
 The values of $\sigma_{in}$ for proton-proton scattering at
 high energies we take from Ref.\cite{GLP}, which describes all
 available data on measured soft cross sections.

We compare with the experimental data of ALICE collaboration
\cite{ALICEPT1,ALICEPT2,ALICEPT3,ALICEPT4} on the tansverse
 momentum distribution  of the charged hadrons in different
 centrality classes at two energies W = 5 TeV and W = 2.78 TeV.
 We add to \eq{MFA} the thermal term (see first term of \eq{SUM}),
 and find that in our model of hadronization    we need this term
 to describe the experimental data. The value of this contribution
 essentially depends on the mass of the gluon jet in the framework
 of our  model.  It turns out that the temperature
$T_{\rm th}$ is proportional to the value of the saturation scale
 in the given centrality class in accord with \eq{THTEM}, but the 
coefficient $c$ turns out  to be  1.5 larger than it is predicted in
 Ref.\cite{KLT} ($ c\, =\,1.8$).

 
\begin{figure}[ht]
\begin{tabular}{c c c}
 \includegraphics[width=5.5cm]{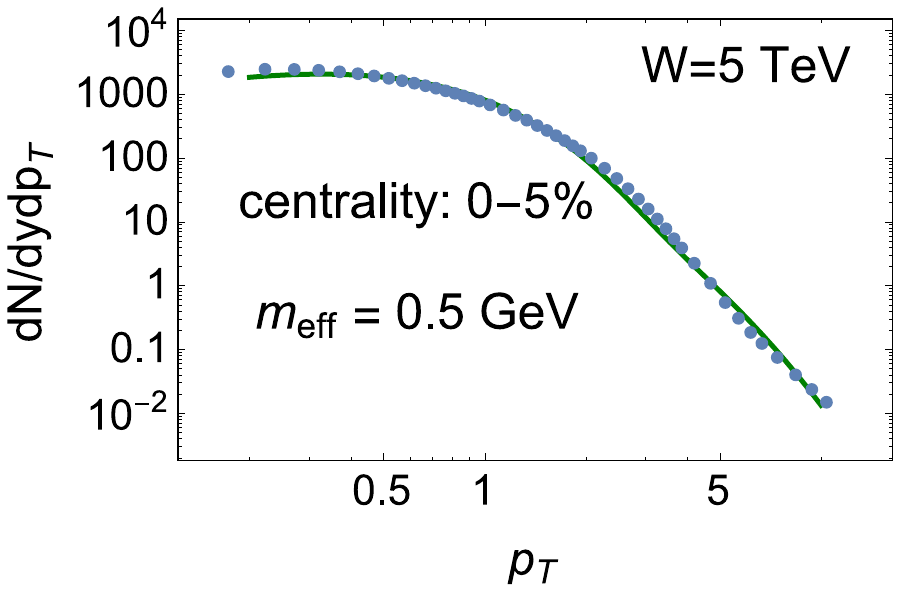}&~~~~~~~~~~~~& \includegraphics[width=5.5cm]{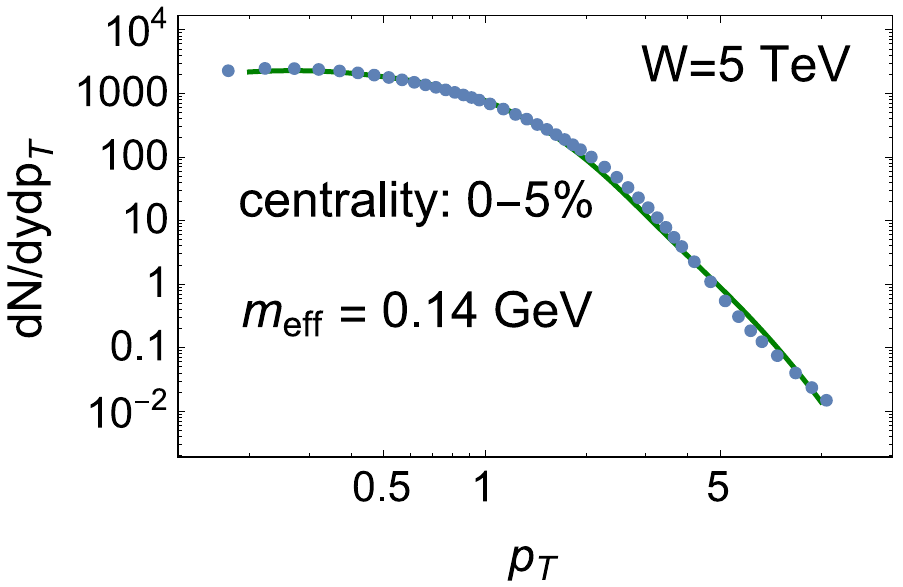}\\
  \includegraphics[width=5.5cm]{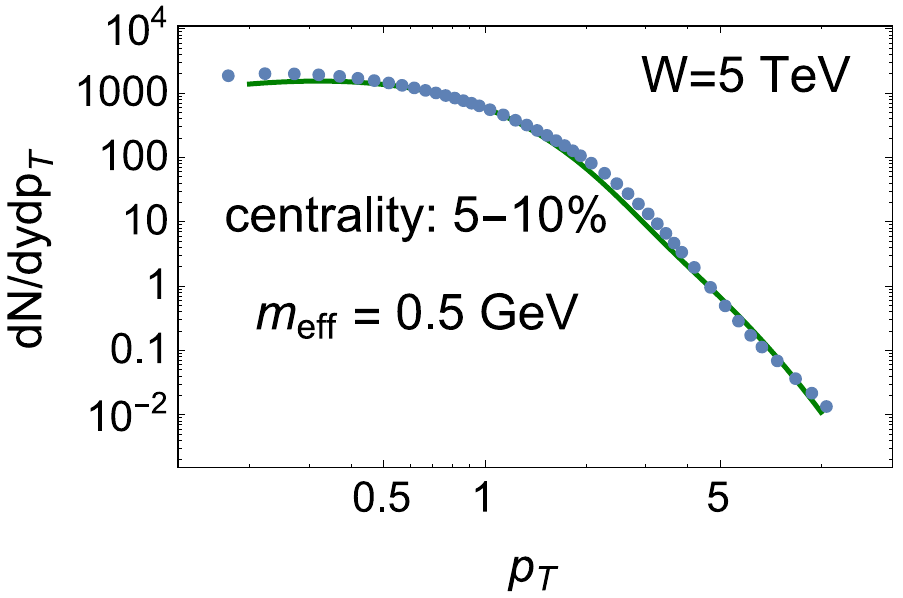}&~~~~& \includegraphics[width=5.5cm]{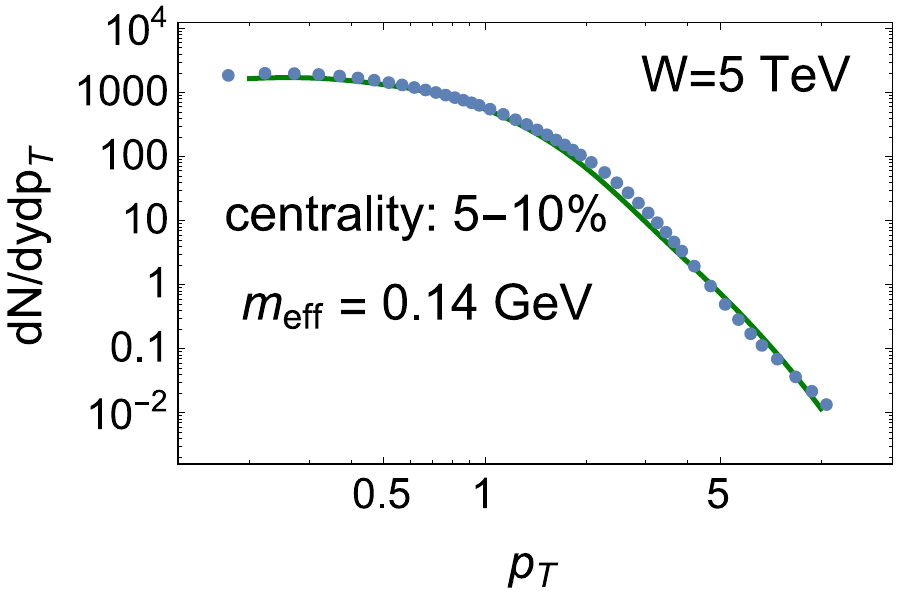}\\
   \includegraphics[width=5.5cm]{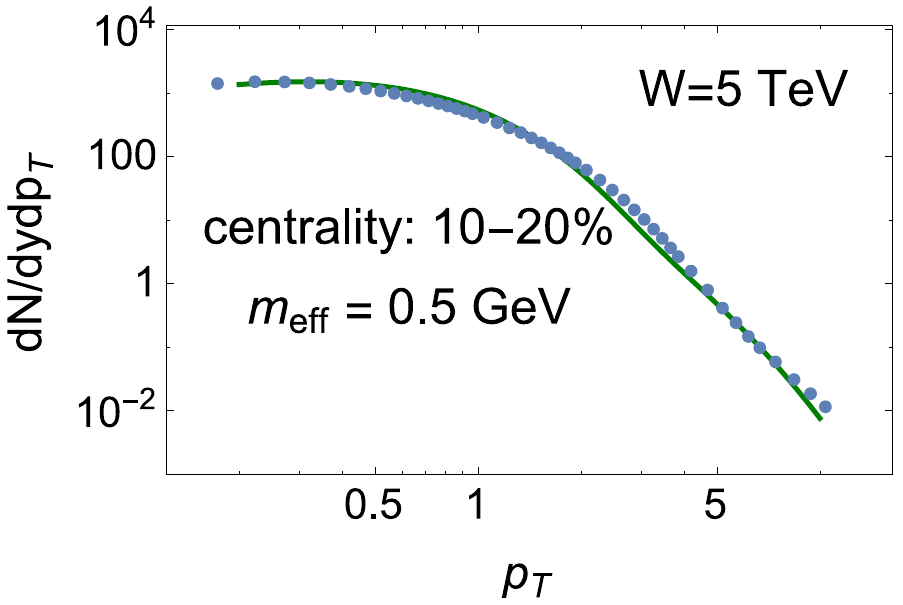}&~~~~& \includegraphics[width=5.5cm]{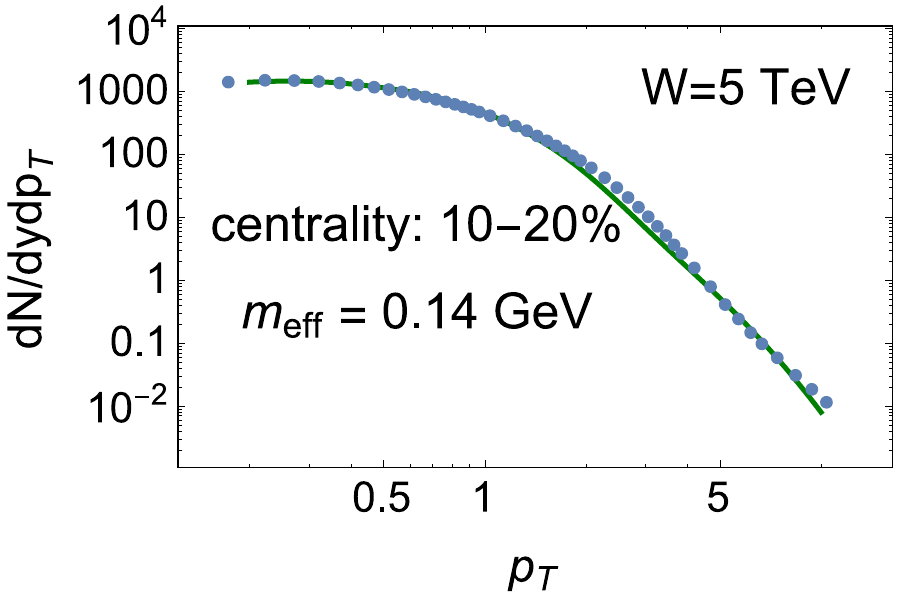}\\ 
            \includegraphics[width=5.5cm]{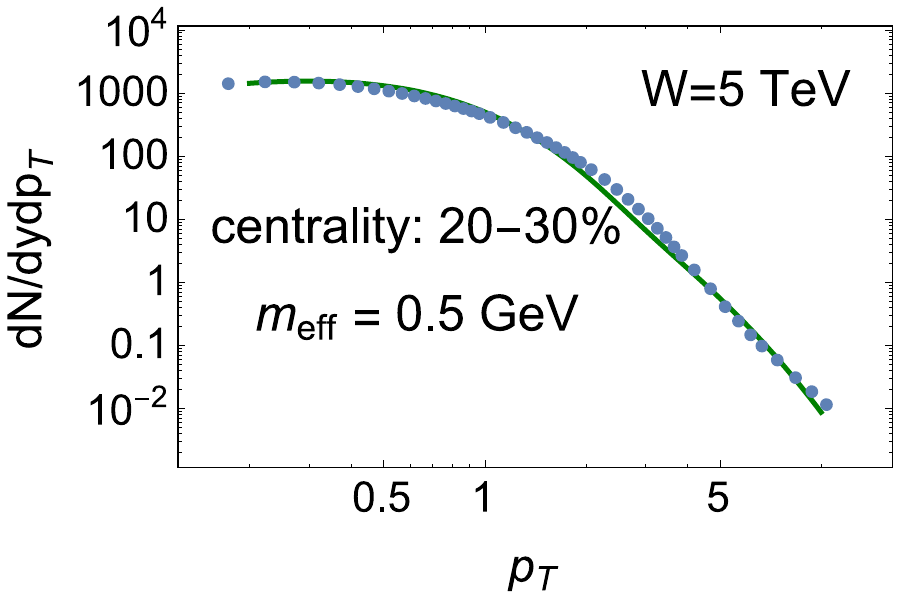}&~~~~& \includegraphics[width=5.5cm]{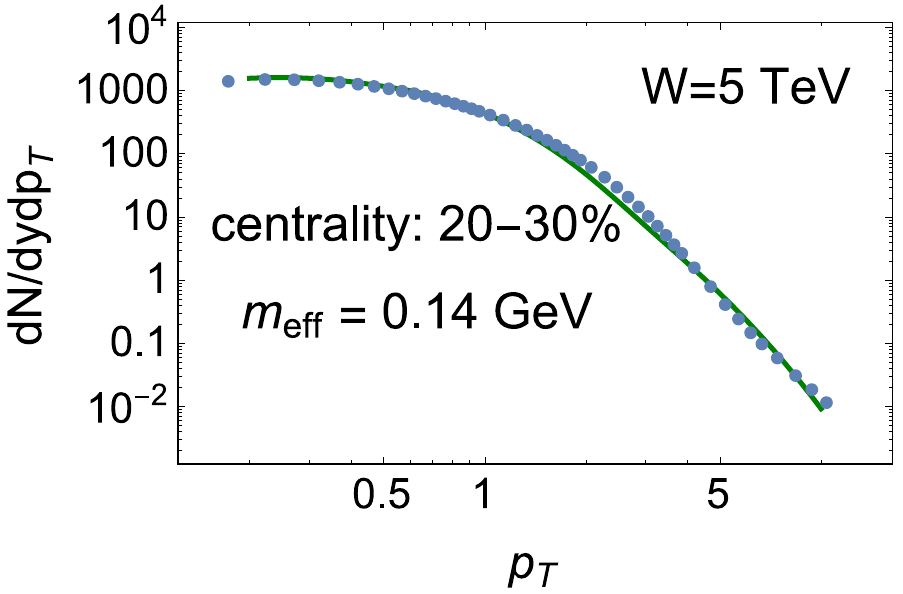}\\  
     \end{tabular}
    \protect\caption { Descriptions of the experimental data
 of the ALICE collaboration\protect\cite{ALICEPT3,ALICEPT4} for lead-lead
 collisions at $W = 5\,TeV$ for different centralities. For the value
 of $\sigma_{in}$ in\eq{MFA} 
 at $W = 5 \,TeV$ we use the model of Ref.\protect\cite{GLP},  for
 $\sigma_{in}\,\,=\,\,\sigma_{tot} - \sigma_{el} - \sigma_{diff}$. 
 }
\label{fit}
   \end{figure}
\begin{figure}[ht]
\begin{tabular}{c c c}  
       \includegraphics[width=5.5cm]{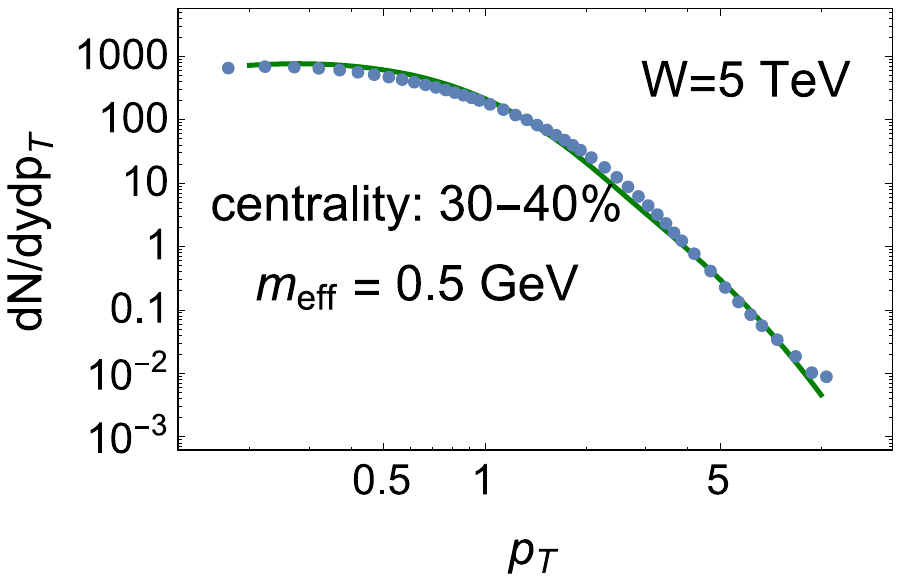}&~~~~~~~~~& \includegraphics[width=5.5cm]{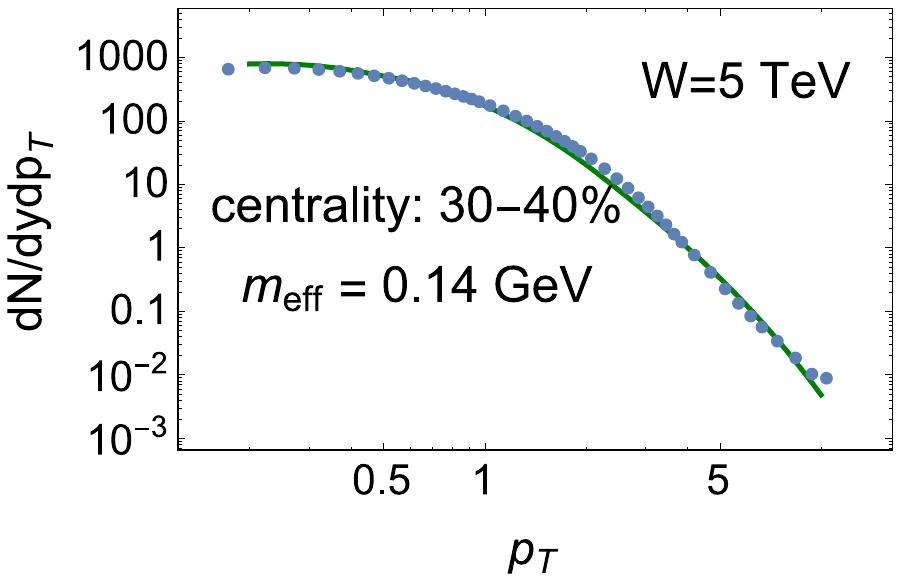}\\         
               \includegraphics[width=6cm]{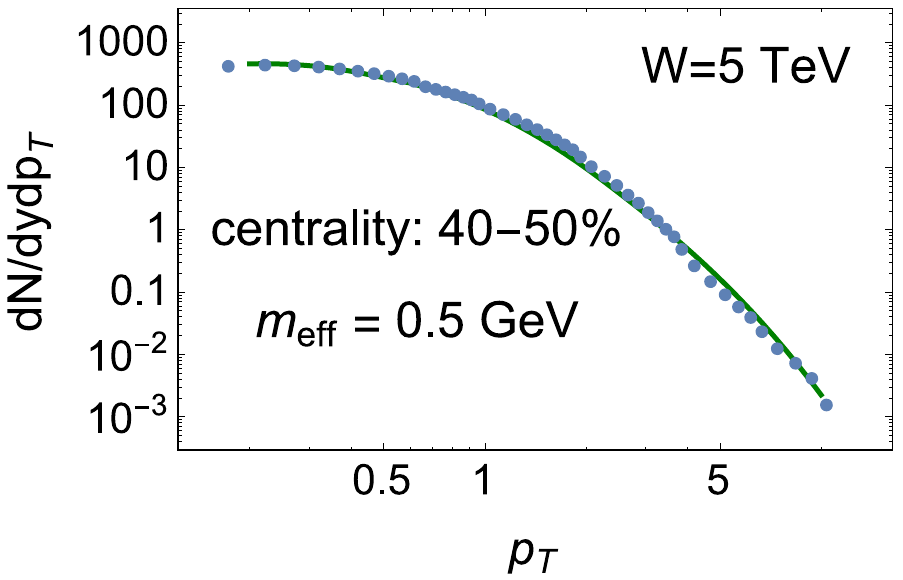}&~~~~& \includegraphics[width=5.5cm]{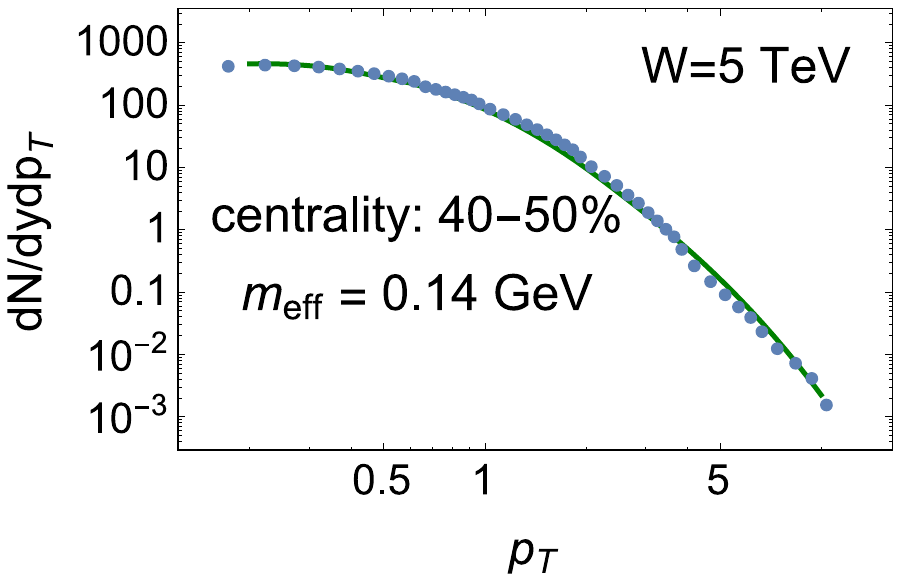}\\             
   \includegraphics[width=5.5cm]{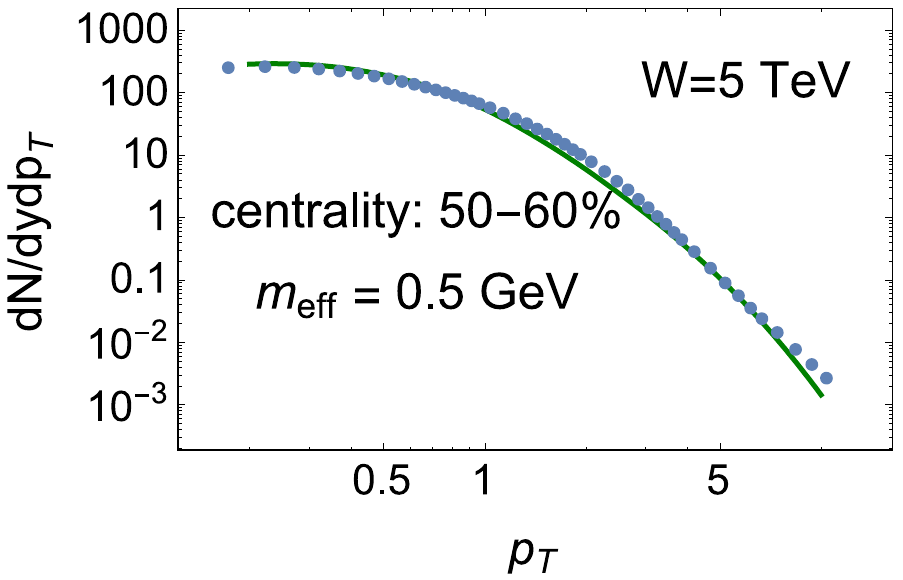}&~~~~& \includegraphics[width=5.5cm]{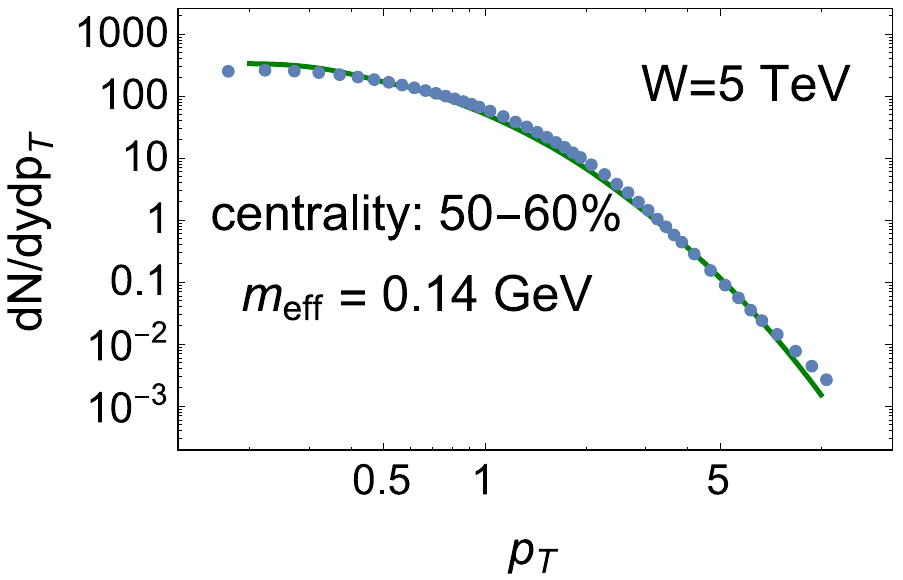}\\ 
   \includegraphics[width=5.5cm]{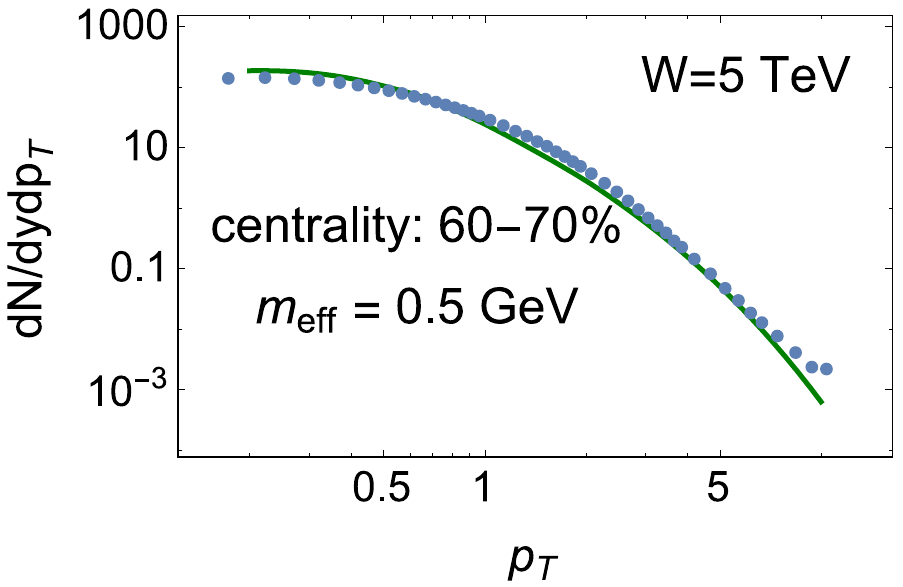}&~~~~& \includegraphics[width=5.5cm]{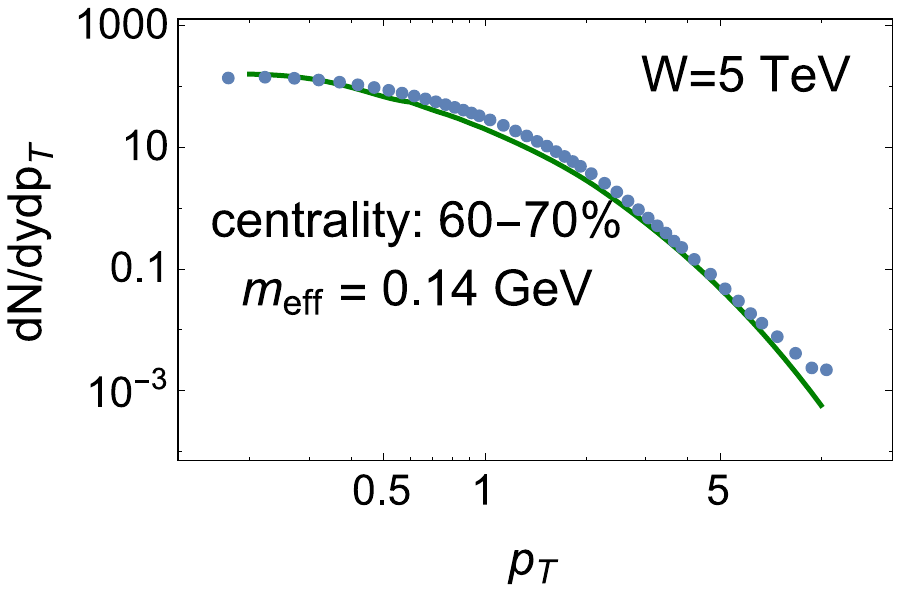}\\ 
 \end{tabular}
    \protect\caption{ The continuation of \fig{fit} 
 }
\label{fit1}
   \end{figure}
\begin{figure}[ht]
\begin{tabular}{c c c}
 \includegraphics[width=5.5cm]{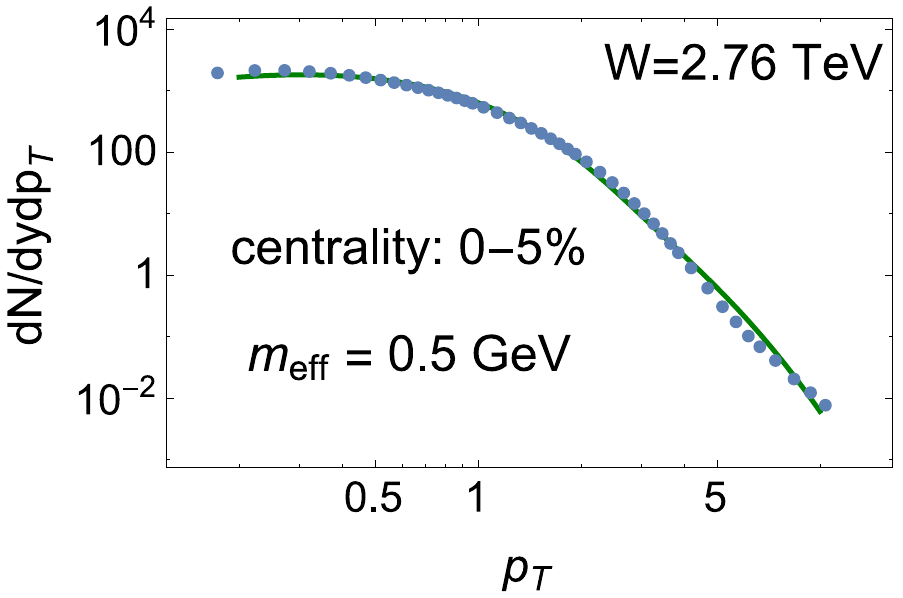}&~~~~~~~~~~~& \includegraphics[width=5.5cm]{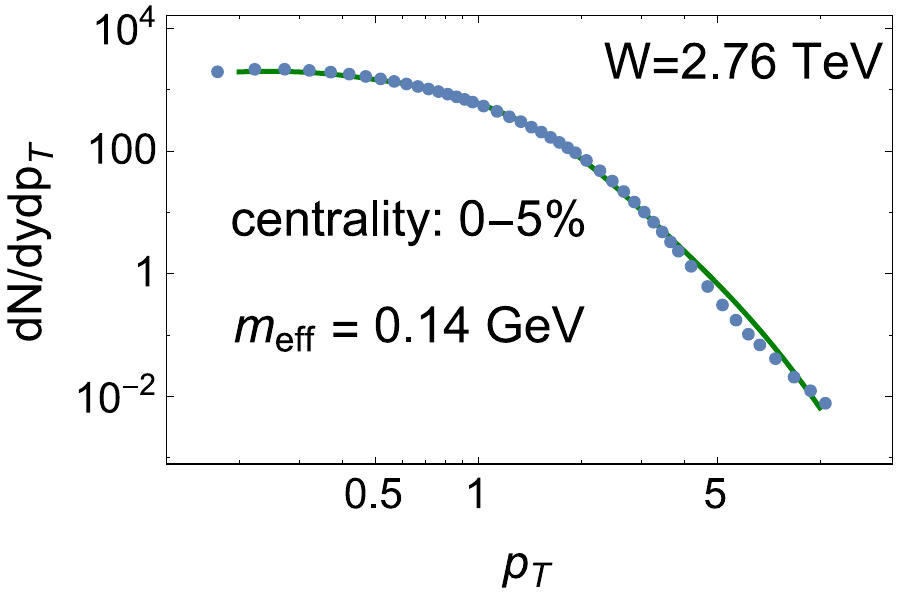}\\
  \includegraphics[width=5.5cm]{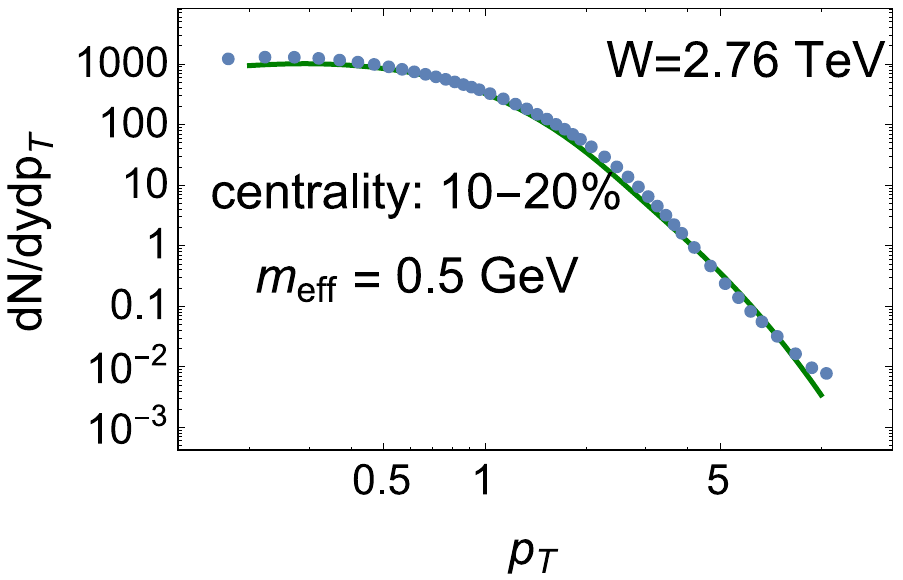}&~~~~& \includegraphics[width=5.5cm]{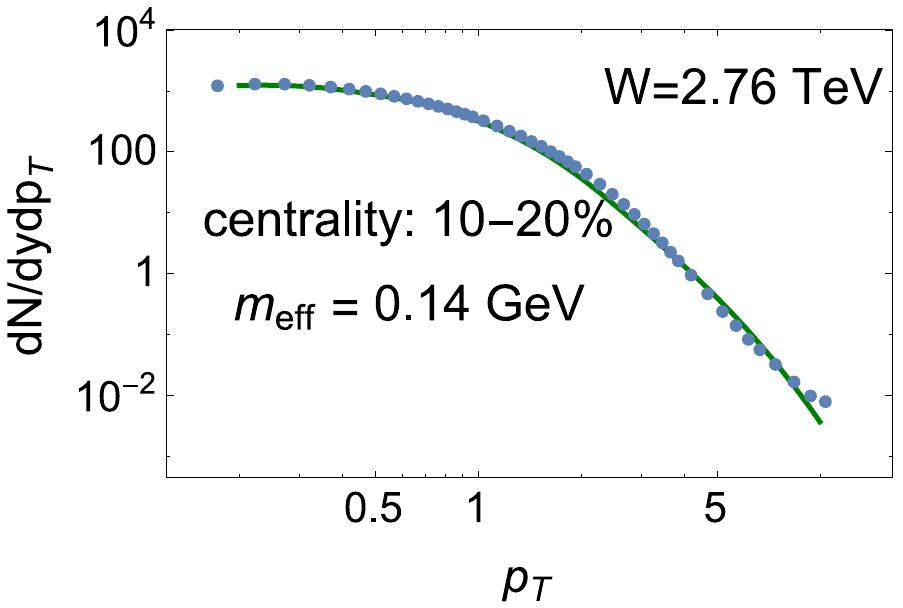}\\
   \includegraphics[width=5.5cm]{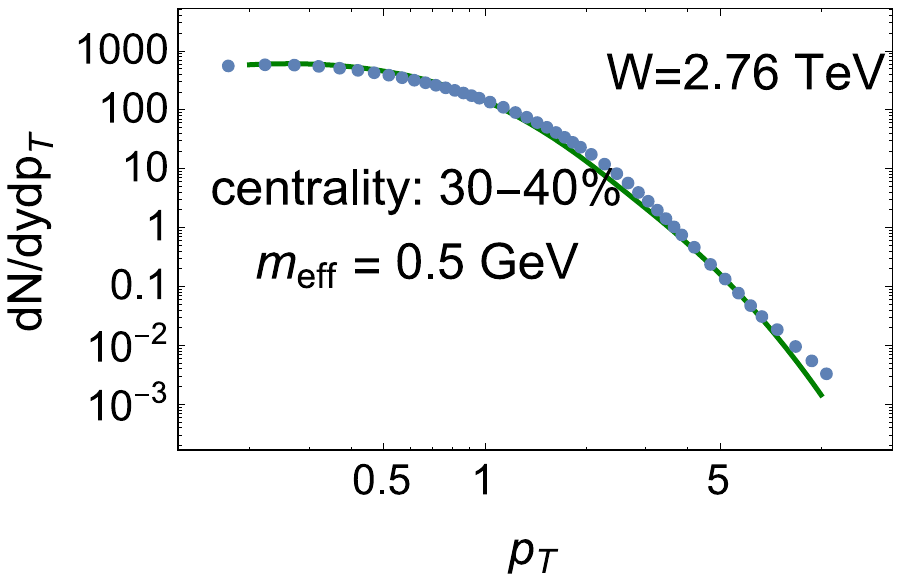}&~~~~& \includegraphics[width=5.5cm]{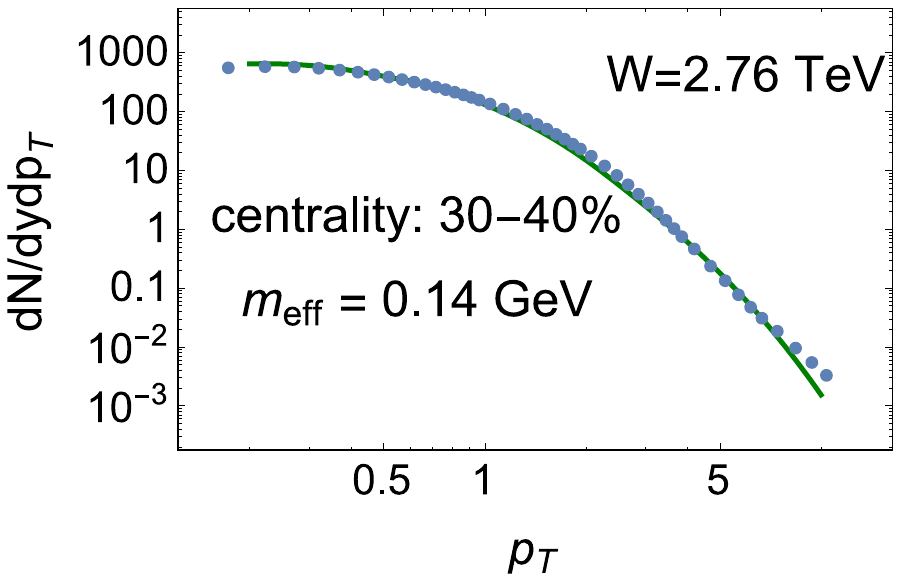}\\ 
 \includegraphics[width=5.5cm]{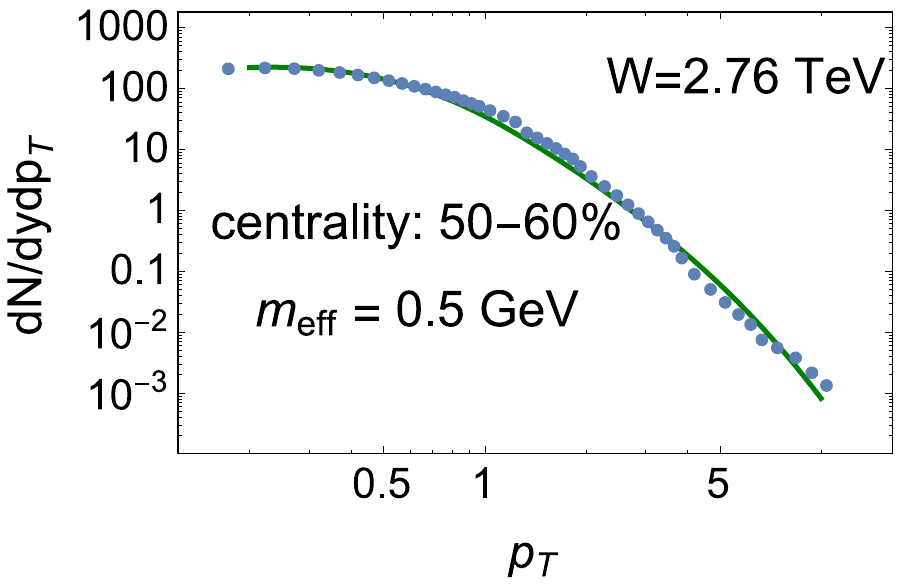}&~~~~& \includegraphics[width=5.5cm]{AADNDPT60053Tev.pdf}\\  
     \end{tabular}
    \protect\caption{ The descriptions of the experimental data
 of the ALICE collaboration\protect\cite{ALICEPT2,ALICEPT3,ALICEPT4} for
 lead-lead collisions at $W = 2.76\,TeV$ for different centralities.
 For the value of $\sigma_{in}$ in    \eq{MFA} 
 at $W = 2.76 \,TeV$ we use the model of Ref.\protect\cite{GLP},  for
 $\sigma_{in}\,\,=\,\,\sigma_{tot} - \sigma_{el} - \sigma_{diff}$. 
 }
\label{fit2}
   \end{figure}

 In \fig{fit}, \fig{fit1} and \fig{fit2} we show that we describe
 the data fairly well in the region $p_T \leq\,10\,GeV$. It should be
 noted that we do not need any $K$-factor which would account for
 the higher order corrections in the framework of CGC/saturation approach. 
However,  this conclusion we need  to take with a  degree of  
scepticism, since the value  of $\sigma_{in}$ was taken from the
 model\cite{GLP}, and the value of $\bas$ in \eq{MFA} was taken
 $\bas =0.25$. The main uncertainty in
 $\sigma_{in}\,=\,\sigma_{tot} - \sigma_{el} - \sigma_{diff}$ 
 is the value of $\sigma_{diff}$, which at high energies is
 about 15 - 20\% of the total cross sections.  The value of
 $\bas\Lb Q^2_s\Rb  \approx 0.3 - 0.4$. Therefore, these two
 uncertainties can lead to the  $ K\,  \approx\, 2$.

 \fig{fit},   \fig{fit1} and \fig{fit2} show that we describe
  the behaviour of the transverse momentum distribution
 at $p_T \,\geq\,2$ by \eq{MFA}, rather well. We wish  to note that 
this
 description stems from the expression for $\gamma_{\rm eff}$ (see
 \eq{GAEFFA} ), in which the last term comes from the corrections
 related to the nucleus target. In the description of $p_T$ distribution,
 these corrections are considerable.

 The rate of  thermal radiation is shown in Table 2, in which
 $R = \int  d^2 p_T \,d^2 \sigma^{\rm charged}_{\mbox{therm. rad.}}
 /d^2 p_T/\int  d^2 p_T \,d^2 \sigma^{\rm charged }_{\mbox{sum.}}
 /d^2 p_T$. Note  that the contribution of the thermal
 radiation increases with the growth of energy. The value of
 the CGC term depends on the value of the $m_{\rm eff}$. We
 believe that  most of the pions are produced from  $\rho$
 resonances and we consider $m_{\rm eff} =  0.5\, GeV  $.
  We recall that the simple formula for $m_{\rm 
 eff}= \sqrt{\mu^2 + k^2_T +k^2_L} - k_L$ leads to
 $m_{\rm  eff} = 0.5\,GeV$  if $\mu$ is equal to the
 mass of $\rho$-resonance since the value of $k_T = k_L
  = 0.45\,GeV$ (see Ref.\cite{ATLASJET} for the measurement
 and Ref.\cite{KLRX}) and reference therein for theoretical
 discussions). For the minimal mass of $\mu = m_\pi = 0.14 \,GeV$
 we obtain $m_{\rm  eff}\,=0.2\,GeV$.  
 To illustrate the influence of the mass of gluon jet we estimate
 the contribution of \eq{MFA} with $m_{\rm eff} = 0.14\,MeV$. One
 can see that even such small mass cannot describe the spectrum
 without the thermal radiation term.
 
\begin{figure}[ht]
\begin{tabular}{c c c}
 \includegraphics[width=7cm]{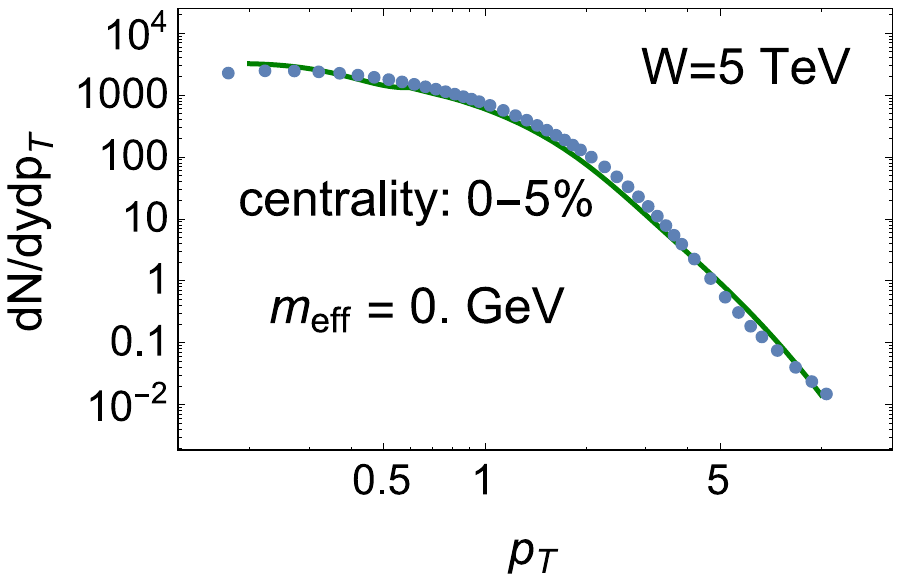}&~~~~~~~& \includegraphics[width=7cm]{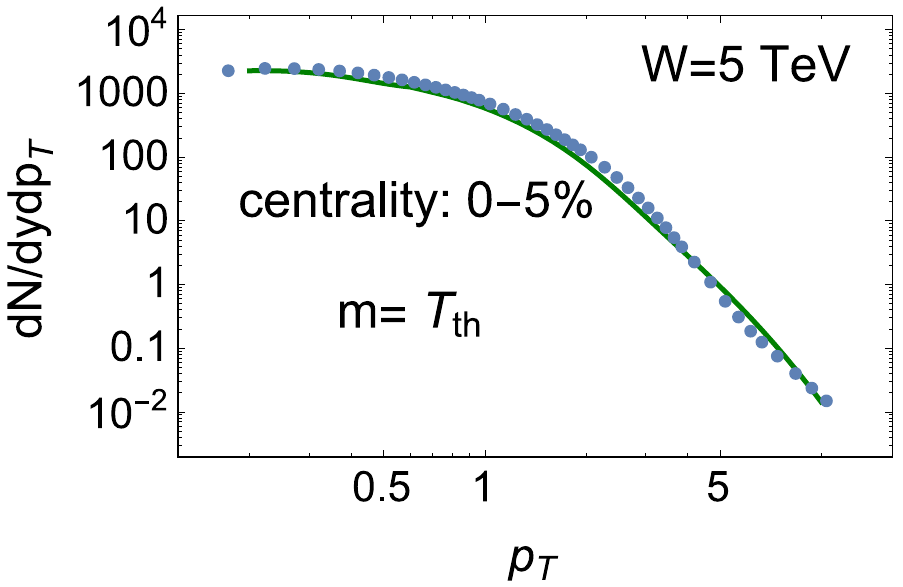}\\
  \fig{m0}-a & & \fig{m0}-b\\
  \end{tabular}
    \protect\caption{  Descriptions of the experimental data
 of the ALICE collaboration\protect\cite{ALICEPT3,ALICEPT4} for lead-lead
 collisions at $W = 5\,TeV$ for centrality 0-5\% with $m_{\rm eff}=0$ (
 \fig{m0}-a) and with $m = T_{\rm th}$ for the gluon propagator
 $1/(p^2_T + m^2)$ (\fig{m0}-b).  In both figures R \,$\approx$\,\,46\%.
 }
\label{m0}
   \end{figure}

 In \fig{m0} we plot the estimates for the $p_T$ distribution with
 $m_{\rm eff}=0$. It turns out that we need to add the thermal
 emission with R = 42\%. This should be compared with the
 proton-proton scattering \cite{GOLETE}, where for $m_{\rm eff}=0$ we
 do not need to add the thermal emission.

\begin{table}[h]
\begin{tabular}{|l|l|l|l|l|}
\hline
  & \multicolumn{2}{c}{ W = 5 TeV} |& \multicolumn{2}{c}{ W = 2.76 TeV} |\\
\hline 
centrality& $m_{\rm eff}$ = 0.5 \,GeV) & $m_{\rm eff}$ = 0.14 \,GeV&$m_{\rm eff}$ = 0.5 \,GeV) & $m_{\rm eff}$ = 0.14 \,GeV  \\
\hline
0-5\% & 76\% &      63\% & 66\%  & 52\%  \\
 \hline
 5-10\% & 72\% &   58\%    &   &   \\ 
 \hline
 10-20\% & 72\% &       58\%  & 52\% & 43\% \\
 \hline
 20 -30\% & 72\% &  54\%   &  &   \\ 
 \hline
 30-40\% & 62\% &      37\% & 62\%  & 34\%  \\ 
 \hline
 40-50\% & 60\% &      37\% &   &   \\
  \hline
 50-60\% &51\% &      20\% & 60\%   & 12\%   \\ 
  \hline
 60-70\% &43\% &      13\% &   &   \\ 
 \hline
  \end{tabular}
\caption{$R\,\,=\,\,d^2\sigma^{\rm charged}/d^2p_T (\mbox{thermal radiation})
\Big{/}
 d^2\sigma^{\rm charged}/d^2p_T (\mbox{sum})$ for different 
 the centrality classes  versus the 
 values of $m_{\rm eff}$. }
\label{t2}
\end{table}

In general, comparing  table II with the estimates for  the ratio R
 for proton-proton scattering (see Table II of Ref.\cite{GOLETE}) we
 see that the contributions of the thermal radiation are larger for
 ion-ion collisions  for the classes with small centrality. It
 supports 
the CGC picture in which  the longitudinal fields, that are the sources
 of the thermal radiation, are stronger in the denser  gluon states 
 which are  produced
 in ion-ion collisions.


Discussing  hadron production, we have to  construct a model for 
the 
 process of hadronization.  Our model where  the production of the gluon
 jets with the hadronization,  which is given by the fragmentation
 functions and by the mass of the gluon jet, requires a 
thermal
 radiation term. Our estimates for $m_{\rm eff}=0$ show that,
possibly, our
 claim does not depend on the model of the hadronization in the case of the
 ion-ion collisions, since the description of the experimental data in  
\fig{m0}-a requires a thermal radiation term of the order of $46\%$.
  In \fig{m0}-b
 we plot the $p_T$ spectrum for a different model, with the gluon
 propagator $1/(p^2_T + m^2)$ with $m = T_{\rm th}$, instead of 
\eq{GLPROP}. 
 It turns out that R\,=\,46\% is needed to describe the data.
 The last case corresponds to a different  
 hadronization model : the propagator of the
 gluon with transverse momentum $p_T$  in the  CGC medium with
 the temperature $T_{\rm th}$,  acquires a mass
 $m_g\,\,\propto\,\, T_{\rm th}$\cite{BRPI} and 
the propagator  has  the form $1/(p^2_T + m^2_g)$.
 This mass provides the infrared cutoff in the gluon
 spectrum.  However, we need to take our estimates in \fig{m0}-b with a
 grain  of salt, since Ref. \cite{BRPI}  predicts that the gluon mass will 
be $m\,=g\,T$ but with small value of $g$. However, \fig{m0}-a supports
 our claim, that the  requirement  of the thermal emission does not 
depend on the
 details of the assumption on the confinement of quarks and gluon.

  \section{Conclusions}

    The main result of the paper is, that we show that a thermal emission
  term is required to describe the transverse momenta distribution for 
charged
 particles in ion-ion collision. The temperature of this emission
 $T_{\rm th}$  turns out to proportional to the saturation scale
 (see \eq{THTEM})  with coefficient $c = 1.8$, which is  1.5 times 
 larger than  predicted in Ref.\cite{KLT}.

  We develop the formalism  for the calculation of the transverse
 momenta spectra in CGC/saturation approach, in which we clearly see two 
stages of the process: the creation of colour glass condensate; and the 
hadronization stage.
 Our calculations  are based on the
 observation that even for small values of $p_T$ the main contribution in
 the integration over $r$ in \eq{MF} and in \eq{MFA} 
 stems from the kinematic region in the vicinity of the saturation 
momentum, 
where theoretically, we know the scattering amplitude. In other words,
 it means that we do not need to 
introduce the non-perturbative corrections due to the unknown physics at long
 distances (see Refs.\cite{VAZW,KHALE} for example) in the dipole scattering
 amplitude. The non-perturbative corrections have to be included to 
describe
 the process of hadronization, which we discuss in the 
 model.  This model incorporates  the decay of the gluon jet with the
 effective mass  $m^2_{\rm eff} = 2 Q_s \mu_{\rm soft}$ where $\mu_{\rm
 soft}$ is the soft scale, and with the fragmentation functions of 
\eq{FRF},
 at all  values of the transverse momenta.

 It should be emphasized that we reproduce the experimental data without
 any K-factor, which is used  for accounting of the higher order
 corrections.  We wish to  mention, that we have calculated the
 inclusive production taking $\bas = 0.25$. This value is less that
 $\bas\Lb Q_s\Rb\,\,=\,\,0.3 - 0.4$ which appears more natural in \eq{MFA}.
 For $\bas = \bas\Lb Q_s\Rb$, we need to introduce a K-factor of about
 1.3 - 1.6.
 
 We use  the KLN model \cite{KLN1,KLN2,KLN3,KLN4,KLN5,DKLN} which  gives us
 the simple way to estimate the cross sections for the different centrality
 classes.
At first sight, we do not need \eq{MFA},  since we can use the master
 equation ( see \eq{MF}), and using the  experimental information of gluon
 structure functions of nucleus (see for example Ref.\cite{AKER})  we can
 calculate the inclusive cross section, as we did for proton-proton 
scattering
 in Ref.\cite{GOLETE}. However, considering the different centrality classes,
  we make the additional selection on the multiplicity of produced hadrons,
 which violates the factorization of \eq{MF}. The  KLN model suggests  a 
way to
 evaluate the contribution of the different centrality classes.

Comparing the results of this paper with our discussion of the transverse
 distribution in the proton-proton scattering\cite{GOLETE} we see two major
 differences. First, we need the larger contribution of the thermal radiation
 term, since we produce in the ion-ion collision the CGC with higher 
parton
 densities. Second, changing the model for the hadronization, 
we failed to describe the $p_T$ spectrum without the
 thermal radiation term.
 In \fig{m0} we illustrate this fact showing that the divergence, 
coming from the factor $1/p^2_T$ in \eq{MFA} , does not
lead to a sufficiently large contribution that we could describe the data 
without
 the thermal radiation term, as  was  the case for hadron-hadron
 scattering \cite{GOLETE}.  Therefore, we suspect 
that
 that the existence of the thermal radiation term does not depend on
 the model of confinement.

  \section{Acknowledgements}
   We thank our colleagues at Tel Aviv university and UTFSM for
 encouraging discussions. Our special thanks go to  
  Keith Baker and Dmitry Kharzeev    for  fruitful 
 discussions on the subject which   prompted the appearance of this 
paper. 
 
  This research was supported  by 
   Proyecto Basal FB 0821(Chile),  Fondecyt (Chile) grant  
 1180118 and by   CONICYT grant PIA ACT1406.

\end{document}